\documentstyle[preprint,aps,epsfig]{revtex}
\begin{document}
%
%
\preprint{$
\begin{array}{l}
\mbox{FERMILAB-Pub-00/153-T}\\[-3mm]
\mbox{UB-HET-00-01}\\[-3mm]
\mbox{August 2000} \\ [1cm]
\end{array}
$}
\title{Probing Neutral Gauge Boson Self-interactions in $ZZ$ Production
at Hadron Colliders\\[1cm]}
\author{U.~Baur\footnote{e-mail: baur@ubhex.physics.buffalo.edu}}
\address{Department of Physics,
State University of New York, Buffalo, NY 14260, USA\\[3.mm]}
\author{D.~Rainwater\footnote{e-mail: dlrain@fnal.gov}}
\address{Theory Group, Fermi National Accelerator Laboratory, Batavia,
IL 60510, USA\\[10.mm]}
\maketitle
%
%
%
\begin{abstract}
\baselineskip15.pt  
A detailed analysis of $ZZ$ production at the upgraded Fermilab Tevatron and 
the CERN Large Hadron Collider is presented for general $ZZZ$ and $ZZ\gamma$ 
couplings. Deviations from the Standard Model gauge theory structure for each 
of these can be parameterized in terms of two form factors which are severely 
restricted by unitarity at high energy. Achievable limits on these couplings 
are shown to be a dramatic improvement over the limits currently obtained by 
$e^+e^-$ experiments.
\end{abstract}
\newpage
\tightenlines
\section{Introduction}
\label{sec:intro}

The Standard Model (SM) of electroweak interactions makes precise predictions 
for the couplings between gauge bosons due to the non-abelian gauge 
symmetry of 
$SU(2)_L\otimes U(1)_Y$. These self-interactions are described by the triple 
gauge boson (trilinear) $WWV$, $Z\gamma V$, and $ZZV$ ($V=\gamma,\,Z$) 
couplings and the quartic couplings. Vector boson pair production provides a 
sensitive ground for {\em direct tests} of the trilinear couplings. Deviations 
of the couplings from the expected values would indicate the presence of new 
physics beyond the SM.

To date the SM has passed this rigorous test with no observed deviations from 
the SM values. The $WWV$ couplings have been measured with an accuracy
of $10-15\%$ in $W^+W^-$, single photon and single $W$ production at
LEP2~\cite{LEP2}, 
and with $20-40\%$ accuracy in $W\gamma$, $WZ$ and $W^+W^-$ production at the
Fermilab Tevatron collider~\cite{cdfwwv,cdfww,d0wwv,d0ww}. The $Z\gamma V$
couplings can be probed in $Z\gamma$ production in $e^+e^-$ and in
hadronic collisions. The LEP2~\cite{LEP2} and Tevatron~\cite{cdfzg,d0zg} 
experiments find the $Z\gamma V$ couplings to be smaller than
$0.05-0.4$, depending on the specific coupling considered. The
$ZZV$ couplings, on the other hand, are only loosely constrained at the
moment through $ZZ$ production at LEP2~\cite{LEP2}. Due to low event
rates after branching ratios, or large backgrounds, $ZZ$ production was not
observed by the Tevatron experiments in Run~I.

In Run~II of the Tevatron which will begin in~2001, an integrated
luminosity of $2-15$~fb$^{-1}$ is
envisioned~\cite{lathuile},  and a sufficient number of $ZZ$ 
events should be available to commence a detailed investigation of the 
$ZZV$ couplings. At the CERN Large Hadron
Collider [(LHC), $pp$ collisions  at $\sqrt{s}=14$~TeV~\cite{LHC}], one
can imagine that the measurement of these couplings reaches the $0.1\%$
level of current precision electroweak data.
In this paper we study the capabilities of future hadron
collider experiments to test the $ZZV$ vertices via $ZZ$ production.
In the past, the reaction $p\,p\hskip-7pt\hbox{$^{^{(\!-\!)}}$} \to ZZ$
for non-zero $ZZV$ couplings has only been studied in the approximation
where the $Z$ bosons are considered as stable final state
particles~\cite{fawzi,glr}. We go a step further and include $Z$ decays
with full decay correlations, finite $Z$ width effects and time-like
virtual photon exchange in our analysis. 

Two $ZZZ$ couplings, and two $ZZ\gamma$ couplings, are allowed by
electromagnetic gauge invariance and Lorentz invariance~\cite{Wcoupling} 
for on-shell $Z$ bosons. We discuss the properties of these couplings 
in Section~\ref{sec:coups}, where we also derive unitarity
bounds for the form factors associated with the $ZZV$ vertices.
The SM is assumed to be valid in our calculations except 
for the $ZZV$ anomalous couplings; $Vf\bar{f}$ couplings and strong
interactions of SM particles remain unchanged.

Our analysis examines the observable final state signatures at 
hadron colliders, $ZZ\to \ell_1^+\ell_1^-\ell_2^+\ell_2^-$, 
$\ell^+\ell^-\nu\bar{\nu}$, $\ell^+\ell^- jj$
($\ell,\,\ell_1,\,\ell_2=e,\,\mu$) and $\bar\nu\nu jj$. 
In Section~\ref{sec:sigs} we provide details of the signal and various 
backgrounds and discuss the signatures of anomalous $ZZZ$ and $ZZ\gamma$ 
couplings. Besides the $ZZ$ invariant mass distribution and the $Z$
boson transverse momentum distributions, the azimuthal angle between the
$Z$ boson decay fermions, $\Delta\Phi$, and their separation 
in the pseudo-rapidity -- azimuthal angle plane, $\Delta R$, are sensitive
indicators of anomalous couplings. The $\Delta\Phi$ distribution may be 
useful in discriminating different types of $ZZV$ couplings, and in
determining their sign. 
In Section~\ref{sec:limits} we derive sensitivity limits for anomalous $ZZV$
couplings for various integrated luminosities at the Tevatron and LHC 
and discuss the results. Finally, in Section~\ref{sec:concl} we 
compare the expectations for Tevatron Run~II and the LHC with the
current LEP2 limits and expectations for an $e^+e^-$ Linear Collider. In 
Section~\ref{sec:concl} we also present our conclusions.

\section{$\boldmath{ZZZ}$ and $\boldmath{ZZ\gamma}$ anomalous couplings}
\label{sec:coups}

\begin{figure}
\phantom{x}
\vskip 8.cm
\includegraphics{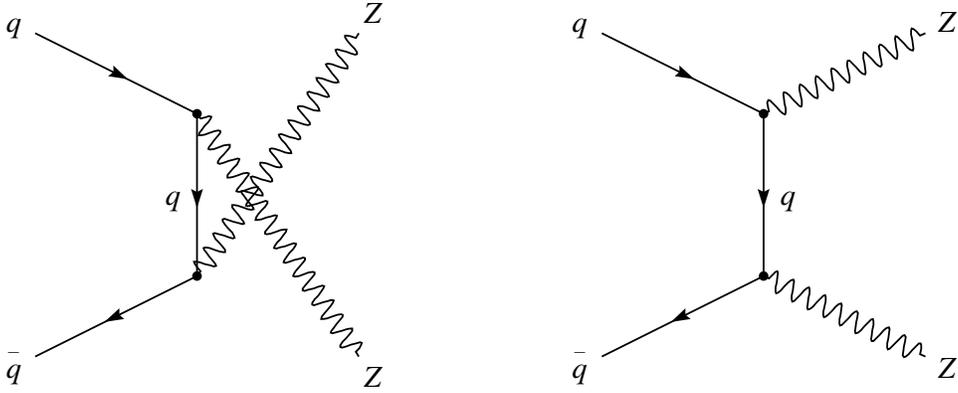}
\caption{The Feynman diagrams for the tree level processes contributing
to $p\,p\hskip-7pt\hbox{$^{^{(\!-\!)}}$} \to ZZ$ in the SM.}
\label{fig:feyn1}
\end{figure}

\begin{figure}
\phantom{x}
\vskip 8.cm
\includegraphics{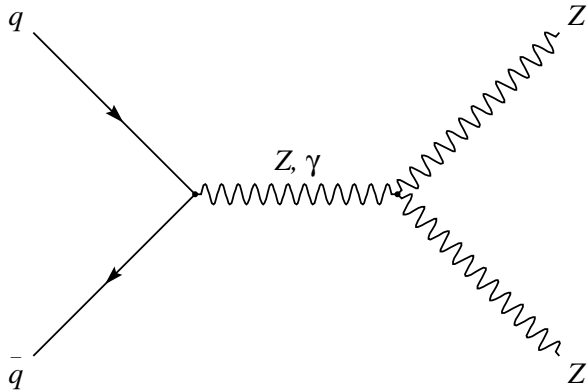}
\caption{Contributions of $ZZZ$ and $ZZ\gamma$ diagrams to $q\bar q\to ZZ$.}
\label{fig:feyn2}
\end{figure}

\begin{figure}
\phantom{x}
\vskip 14.cm
\includegraphics{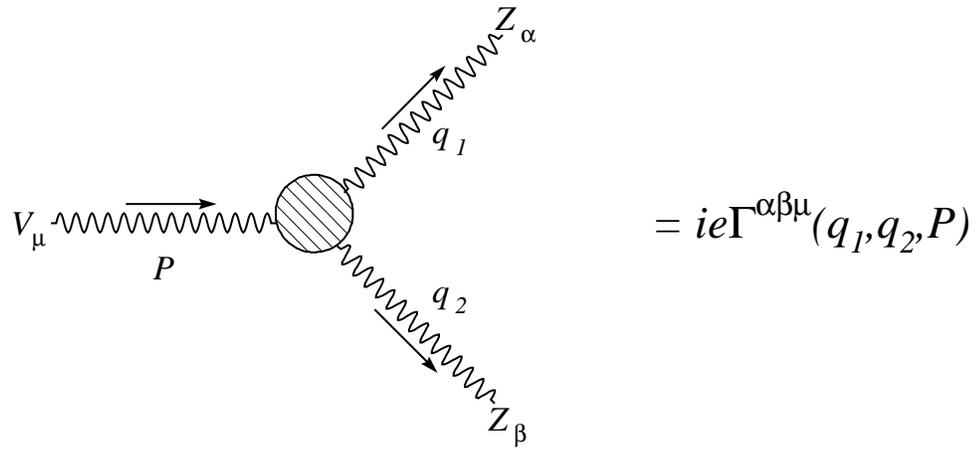}
\caption{Feynman rule for the general $ZZV$ ($V=Z,\,\gamma$) vertex. The 
vertex function is given in Eq.~(\ref{eq:VVV}). $e$ is the charge of the 
proton.}
\label{fig:feyn3}
\end{figure}

In the SM, at the parton level, the reaction
$p\,p\hskip-7pt\hbox{$^{^{(\!-\!)}}$}\to ZZ$ proceeds by the Feynman 
diagrams of Fig.~\ref{fig:feyn1}. Including 
the anomalous couplings under discussion requires the addition of the
graphs of Fig.~\ref{fig:feyn2}. In 
the massless fermion limit, a reasonable approximation for hadron 
collider processes, the most general form of the 
$Z^\alpha(q_1)\,Z^\beta(q_2)\, V^\mu(P)$ ($V=Z,\,\gamma$) vertex
function (see Fig.~\ref{fig:feyn3} for notation) for on-shell $Z$'s
which respects Lorentz invariance and electromagnetic gauge invariance 
may be written as~\cite{Wcoupling} 
\begin{equation}
g_{ZZV} \Gamma^{\alpha\beta\mu}_{ZZV} = 
e\, {P^2-M_V^2\over M_Z^2}\, \biggl[ if_4^V \left(P^\alpha
g^{\mu\beta}+P^\beta g^{\mu\alpha} \right) 
+if_5^V\epsilon^{\mu\alpha\beta\rho}\left(q_1-q_2\right)_\rho\biggr],
\label{eq:VVV}
\end{equation}
where $M_Z$ is the $Z$-boson mass and $e$ is the proton charge. The 
overall factor $(P^2-M_V^2)$ in
Eq.~(\ref{eq:VVV}) is a consequence of Bose symmetry for $ZZZ$
couplings, while it is due to electromagnetic gauge invariance for the
$ZZ\gamma$ couplings. The couplings $f_i^V$ ($i=4,\,5$) are
dimensionless complex functions of $q_1^2$, $q_2^2$ and $P^2$. All
couplings are $C$ odd; $CP$ invariance forbids $f^V_4$ and parity 
conservation requires that $f_5^V$ vanishes. Because $f_4^Z$ and
$f_4^\gamma$ are 
$CP$-odd, contributions to the helicity amplitudes proportional to these
couplings will not interfere with the SM terms. In the static limit, 
$f_5^\gamma$ corresponds to the anapole moment of the $Z$ 
boson~\cite{anapole}. 
In the SM, at tree level, $f^V_4 = f^V_5 = 0$. At the one-loop level,
only the $CP$ conserving couplings $f_5^V$ receive
contributions. Numerically, these contributions are of ${\cal
O}(10^{-4})$~\cite{renard1}. Loop contributions from supersymmetric
particles and additional heavy fermions produce $ZZV$ couplings of a
similar magnitude~\cite{renard1}. If the $Z$ bosons are allowed to be
off-shell, five additional $ZZZ$ couplings, and five additional
$ZZ\gamma$ couplings contribute~\cite{renard2}. 
For these couplings, the factor $(P^2-M_V^2)$ in the vertex function is
replaced by $(q_1^2-q_2^2)$. The effect of these couplings thus is strongly
suppressed and we shall ignore them in our discussion. 

It should be noted that the two $ZZ\gamma$ couplings contributing 
to $ZZ$ production are completely independent of the four $Z\gamma Z$
couplings which appear in $Z\gamma$ production (assuming that the
$Z$-boson is on-shell). If all three vector
bosons in the vertex function are off-shell, there are seven couplings
altogether. Four of them survive in $Z\gamma$ production, and two in
$f\bar f\to ZZ$. 

The parton level di-boson production cross sections with non-SM
couplings manifestly grow  
with the parton center of mass energy $\sqrt{\hat{s}}$. $S$-matrix
unitarity restricts the $ZZV$ couplings uniquely to their SM values at
asymptotically high energies~\cite{unitarity}. This requires that the 
couplings $f^V_i$ possess a momentum dependence which ensures
that the $f^V_i(q^2_1,q^2_2,P^2)$ vanish for any 
momenta much larger than $M_Z$. For $ZZ$ production, $q^2_1,q^2_2 \sim 
M^2_Z$ even considering finite $Z$ width effects, but $P^2 = \hat{s}$ 
may be quite large 
at the hadron colliders under consideration. In order to avoid
unphysical results that would violate unitarity, the $\hat{s}$
dependence thus has to be taken into account. To parameterize the $\hat
s$ dependence of the form factor, 
we will use a generalized dipole form factor~\cite{Zcoupling},
\begin{equation}
f^V_i(\hat{s}) = 
{f^V_{i0} \over ( 1 + \hat{s}/\Lambda_{FF}^{2})^{n} } \quad 
(i = 4,5) \, ,
\label{eq:ff}
\end{equation}
where $\Lambda_{FF}$ is the form factor scale which is related to the
scale of the new physics which is generating the anomalous $ZZV$
couplings. 

The values $f_{i0}^V=f_i^V(M_Z^2,M_Z^2,0)$ of the form factors at low
energy ($\hat s=0$) and the power of the form factor, $n$, are
constrained by partial wave unitarity of the inelastic $ZZ$ production
amplitude in fermion antifermion annihilation at arbitrary
center-of-mass energies. In deriving unitarity limits for the
$f_{i0}^V$'s, we follow the strategy employed in
Ref.~\cite{BZ_PLB201}. The anomalous contribution to the 
\begin{equation}
f(\sigma)\bar{f}(\bar\sigma) \to Z(\lambda_1)Z(\lambda_2)
\end{equation}
helicity amplitudes may be written as 
\begin{equation}
\Delta {\cal M}^V (\sigma\bar{\sigma},\lambda_1\lambda_2)
= -\sqrt{2} \: e^2 g^{V f_1 f_2}_{2\sigma}\, {\hat{s}\over M^2_Z} \,
  \beta \, \delta_{\sigma,\bar{\sigma}} 
  \, A^V_{\lambda_1\lambda_2}
  \times d^1_{\sigma +\bar{\sigma},\lambda_1 -\lambda_2}(\Theta) \; ,
\end{equation}
where $d^1$ are the conventional $d$-functions~\cite{pdg}, $\beta = 
(1-4M^2_Z/\hat s)^{1/2}$, $\sigma,\bar\sigma$ and
$\lambda_1,\lambda_2$ are the helicities of the 
incoming fermion pair and outgoing $Z$ pair, respectively. 
$g^{V f_1 f_2}_{2\sigma}$ is the coupling of the $s$-channel vector
boson to the 
incoming fermion pair, $\Theta$ is the center of mass scattering angle, 
and the $A^V$'s are the reduced amplitudes given by (see also 
Ref.~\cite{chang})
\begin{eqnarray}
A^V_{0 \pm} & = & {\sqrt{\hat{s}}\over 2M_Z} 
            \left [ -if_4^V \pm \beta f_5^V \right ], \label{eq:A1}\\[2.mm] 
A^V_{\pm 0} &= & {\sqrt{\hat{s}}\over 2M_Z} 
            \left [\phantom{-}if_4^V \mp \beta f_5^V \right ],
\label{eq:A2}\\[2.mm] 
A^V_{\pm\pm} & = & A^V_{00} = A^V_{\pm\mp} = 0.
\label{eq:A}
\end{eqnarray}
Examination of the $J=1$ partial wave amplitude produces the desired unitarity 
bounds,
\begin{eqnarray}
\biggl( \sum\limits_{\lambda_1\lambda_2} 
        |A^\gamma_{\lambda_1\lambda_2}|^2 \biggr)^{1/2}
&\leq & {1 \over\alpha\beta^{3/2}} 
      \, \left[{3\over
5}\left(3-6\sin^2{\theta_W}+8\sin^4{\theta_W}\right)\right]^{1/2} 
      \, {M^2_Z \over \hat{s}},  \\[2.mm]
\biggl( \sum\limits_{\lambda_1\lambda_2} 
        |A^Z_{\lambda_1\lambda_2}|^2 \biggr)^{1/2}
&\leq & {4 \over \alpha\beta^{3/2}}\, \sqrt{3\over 10} \,
      \: \sin{\theta_W}\cos{\theta_W} \, {M^2_Z \over \hat{s}},
\label{eq:bounds}
\end{eqnarray}
where $\theta_W$ is the weak mixing angle and $\alpha$ the QED fine 
structure constant.

By substituting Eq.~(\ref{eq:ff}) and assuming that only one coupling is 
nonzero at a time, we find the following unitarity bounds for $\Lambda_{FF}\gg
M_Z$:
\begin{eqnarray}
| f^\gamma_{40,50} | 
&\leq & \; {1\over\alpha}\,\left[{3\over
5}\left(3-6\sin^2{\theta_W}+8\sin^4{\theta_W}\right)\right]^{1/2}\,
\left({M_Z\over\Lambda_{FF}}\right)^3\,{\left({2\over
3}\,n\right)^n\over\left({2\over 3}\,n-1\right)^{(n-3/2)}}\, , 
\label{eq:unilim1} \\
| f^Z_{40,50} | 
&\leq & \; {4\over\alpha}\,\sqrt{3\over 10}\,\sin\theta_W\cos\theta_W\,
\left({M_Z\over\Lambda_{FF}}\right)^3\,{\left({2\over
3}\,n\right)^n\over\left({2\over 3}\,n-1\right)^{(n-3/2)}}\, .
\label{eq:unilim2}
\end{eqnarray}
Tree level unitarity is satisfied throughout the entire $\hat s$ range
when these limits are observed. For the more likely case that
several anomalous couplings contribute, cancellations may occur and the
bounds are weaker than those listed in Eqs.~(\ref{eq:unilim1})
and~(\ref{eq:unilim2}). From the $n$ dependent factors in 
(\ref{eq:unilim1}) and~(\ref{eq:unilim2}) one observes that
$n>3/2$ is necessary in order to satisfy
unitarity. This is a direct consequence of the high energy behavior of
the anomalous contributions to the $ZZ$ helicity amplitudes, which
grow like $(\sqrt{\hat s}/M_Z)^3$. In the following we shall assume that 
$n=3$. Selecting an exponent sufficiently above the minimum value of
$3/2$ ensures that the $ZZ$ differential cross section stays well below
the unitarity limit at energies $\sqrt{\hat s}\gg\Lambda_{FF}\gg M_Z$,
where novel phenomena such as resonance production, or multiple  weak
boson production, are expected to dominate. For the form factor scale we 
choose $\Lambda_{FF}=750$~GeV at the Tevatron and $\Lambda_{FF}=2$~TeV
at the LHC in our numerical simulations.


\section{Signatures of Anomalous $ZZV$ Couplings}
\label{sec:sigs}

In this section we discuss characteristics of the signal and backgrounds of 
anomalous $ZZV$ couplings. For simplicity, we consider only real $ZZV$ 
couplings. We consider four signatures of $ZZ$ production: decays to four 
leptons; two leptons and missing energy; two leptons and two jets; and
two jets plus missing energy. Due to the overwhelming four jet QCD
background, decays where both $Z$ bosons decay hadronically are not
considered here. We calculate the SM signal, the signal with anomalous 
couplings, and the significant backgrounds via full tree level matrix 
elements for the subprocess in question, each of which is discussed in 
detail below. 


\subsection{General considerations}
\label{sec:threeone}

Our calculation is carried out at the tree level. We compute the $q\bar 
q\to ZZ\to 4$~fermion helicity 
amplitudes in the double pole approximation which ignores contributions
from non-resonant diagrams except for contributions from time-like photon 
exchange diagrams, using the method described in Ref.~\cite{hag}. 
Decay correlations, finite $Z$ width effects 
and contributions from time-like photon exchange are taken into
account. Cross sections and dynamical distributions are evaluated using
a parton level Monte Carlo program. 

To simulate the effect of next-to-leading-log (NLL) QCD corrections we
multiply the differential cross section with a simple $K$-factor which
depends on the final state considered. A more detailed discussion of how 
NLL QCD corrections affect the four different final states is presented
in the following sections, where we discuss each final state in turn. 
Gluon fusion, $gg\to ZZ$, contributes about 
$1\%$ ($15\%$) to the cross section at the Tevatron (LHC)~\cite{gb}. We
do not include the contribution from gluon fusion into our analysis. 

To examine the effects of anomalous couplings on observables we simulate 
$p\bar{p}$ ($pp$) collisions at $\sqrt{s}=2$~TeV ($\sqrt{s}=14$~TeV) for 
the Tevatron 
(LHC). For all our numerical results we have chosen the set of SM input 
parameters to be: $\sin^2\theta_W = 0.2310$, $M_Z = 91.187$~GeV, and 
$\alpha(M_Z) = 1/128.93$~\cite{lepsum}. For 
all processes which depend on the QCD coupling constant, we choose the
value of the strong coupling constant to be 
$\alpha_s(M_Z) = 0.118$. We employ CTEQ4L parton distribution 
functions~\cite{CTEQ} for all calculations, selecting the value of the 
factorization scale to be $\mu_f = M_Z$. 

As finite detector resolution can have a sizable effect on cross sections and 
thus the number of events accepted into the data set, to make our calculations 
realistic we must take into account some minimal detector response. We 
accomplish this via Gaussian smearing of the four momenta of the outgoing 
particles according to detector expectations.
For Gaussian smearing at the Tevatron we use the expected values of the 
upgraded CDF detector~\cite{CDF}:
\begin{eqnarray}
{\Delta{E} \over E}{\rm (had)} 
&=& {0.75 \over \sqrt{E_T}} \, \oplus \: 0.03 \quad  (|\eta| < 1.1)
\nonumber \\[2.mm] 
&=& {0.80 \over \sqrt{E}}   \,\,\, \oplus \: 0.05 \quad  (|\eta| > 1.1)
\nonumber \\[2.mm]
{\Delta{E} \over E}{\rm (lep)} 
&=& {0.14 \over \sqrt{E_T}} \, \oplus \: 0.02 \quad  (|\eta| < 1.1)
\nonumber \\[2.mm] 
&=& {0.16 \over \sqrt{E}}   \,\,\, \oplus \: 0.01 \quad  (|\eta| > 1.1) 
\nonumber 
\end{eqnarray}
For the LHC we take the expected values for the ATLAS
detector~\cite{CMS-ATLAS}:
\begin{eqnarray}
{\Delta{E} \over E}{\rm (had)} 
&=& \,\,{0.5   \over \sqrt{E}} \, ~ \oplus \: 0.03  \quad ~
(|\eta| < 3.0) 
\nonumber \\[2.mm] 
{\Delta{E} \over E}{\rm (lep)} 
&=& {0.095 \over \sqrt{E}} \, \oplus \: 0.005 \quad  (|\eta| < 2.5) \nonumber
\end{eqnarray}
Here, $\eta$ is the pseudo-rapidity, $E$ ($E_T$) is the energy 
(transverse energy) measured in GeV, and 
the $\oplus$ sign symbolizes that the two terms are added in quadrature.

In all cases, the missing momentum in an event is taken as the negative vector 
sum of the smeared four momenta of all observable final state particles. This 
does ignore the effects of additional soft activity that will affect this 
distribution in experiment, but for our purposes here may be safely neglected.

The geometric and kinematic acceptance of detectors, i.e. the
ability to observe and properly identify final states particles, 
is simulated in our
calculations by cuts imposed on observable particles in the final state.
At the Tevatron (LHC) we require ($\ell=e,\,\mu$):
\begin{quasitable}
\label{eq:basecuts}
\begin{tabular}{cc}
Tevatron & LHC\\
\tableline
$p_T(\ell) >15$~GeV & $p_T(\ell) >15$~GeV\\
$|\eta(\ell)| <2.5$ & $|\eta(\ell)| <2.5$ \\
$p_T(j)>20$~GeV & $p_T(j)>30$~GeV \\
$|\eta(j)| <2.5$ & $|\eta(j)| <3$\\
$\Delta R(\ell j) >0.6$ & $\Delta R(\ell j) >0.6$\\
$\Delta R(jj) >0.6$ & $\Delta R(jj) >0.6$ \\
$p\llap/_T>20$~GeV for $ZZ\to\ell^+\ell^-\bar\nu\nu$ &
$p\llap/_T>50$~GeV for $ZZ\to\ell^+\ell^-\bar\nu\nu$ \\
$p\llap/_T>60$~GeV for $ZZ\to\bar\nu\nu jj$ & $p\llap/_T>60$~GeV for
$ZZ\to\bar\nu\nu jj$ \\
\end{tabular}
\end{quasitable}
where $p_T$ is the transverse momentum and 
\begin{equation}
\Delta R =
\left[\left(\Delta\Phi\right)^2+\left(\Delta\eta\right)^2\right]^{1/2} 
\end{equation}
is the separation in the pseudorapidity -- azimuthal angle
plane. $p\llap/_T$ is the missing transverse momentum resulting from the 
nonobservation of the neutrino pair.

In addition, we require the $\ell^+\ell^-$ and two jet invariant masses
to be within $\pm 15$~GeV of the $Z$ boson mass~\cite{cdfz,d0z}:
\begin{eqnarray}
76~{\rm GeV} & < m(\ell^+\ell^-) & < 106~{\rm GeV}, \label{eq:mass1}\\
76~{\rm GeV} & < m(jj) & < 106~{\rm GeV}.
\label{eq:mass2}
\end{eqnarray}
These cuts help suppress contributions from non-resonant Feynman
diagrams and, in the jet case, the background from $Z+2$~jet
production. Additional cuts which are imposed to reduce backgrounds for 
individual final states will be discussed when we consider the specific
final state to which they apply. 


\subsection{The {$\mathbf \ell_1^+\ell_1^-\ell_2^+\ell_2^-$} channel}

The first $ZZ$ decay channel we consider,
$ZZ\to \ell_1^+\ell_1^-\ell_2^+\ell_2^-$ ($\ell_1,\,\ell_2 = e,\mu$) is 
observationally the  
cleanest as it is essentially background-free. However, it does suffer from 
a small event rate due to a tiny branching ratio of $B(ZZ\to
\ell_1^+\ell_1^-\ell_2^+\ell_2^-)=0.0045$ if both electron and muon final 
states are considered. In the following we concentrate on the $ZZ\to
e^+e^-\mu^+\mu^-$ channel. Results for the $e^+e^-e^+e^-$ and
$\mu^+\mu^-\mu^+\mu^-$ final states can be obtained by dividing all
cross section results by two. However, for these final states one has to 
take into account the combinatorial background originating from not
being able experimentally to distinguish identical charged leptons. 
All results presented in this section include a 
$K$-factor of 1.28 (1.34) at Tevatron (LHC) energies to approximate the 
effect of NLL QCD corrections~\cite{zznlo}. Inclusive NLL QCD corrections 
to $ZZ$ production are known to modify 
the shape of distributions only insignificantly. 
As mentioned in Sec.~\ref{sec:threeone}, 
we do not include the non-resonant Feynman diagrams which contribute to
$q\bar q\to e^+e^-\mu^+\mu^-$ in our calculation. Requiring the
invariant mass of the lepton pairs to be in the vicinity of the $Z$ mass,
(see Eq.~(\ref{eq:mass1})), these diagrams contribute less than a few
percent to the differential cross section. Imposing the cuts listed in 
Sec.~\ref{sec:threeone}, one obtains a $ZZ\to \ell^+\ell^-\ell^+\ell^-$
($\ell = e,\mu$) cross section of 3.85~fb (22.5~fb) at the Tevatron
(LHC). For 2~fb$^{-1}$ in Run~II, only a few $ZZ\to 4$~lepton events are 
therefore expected within the framework of the SM. 

\begin{figure}
\phantom{x}
\vskip 15cm
\includegraphics{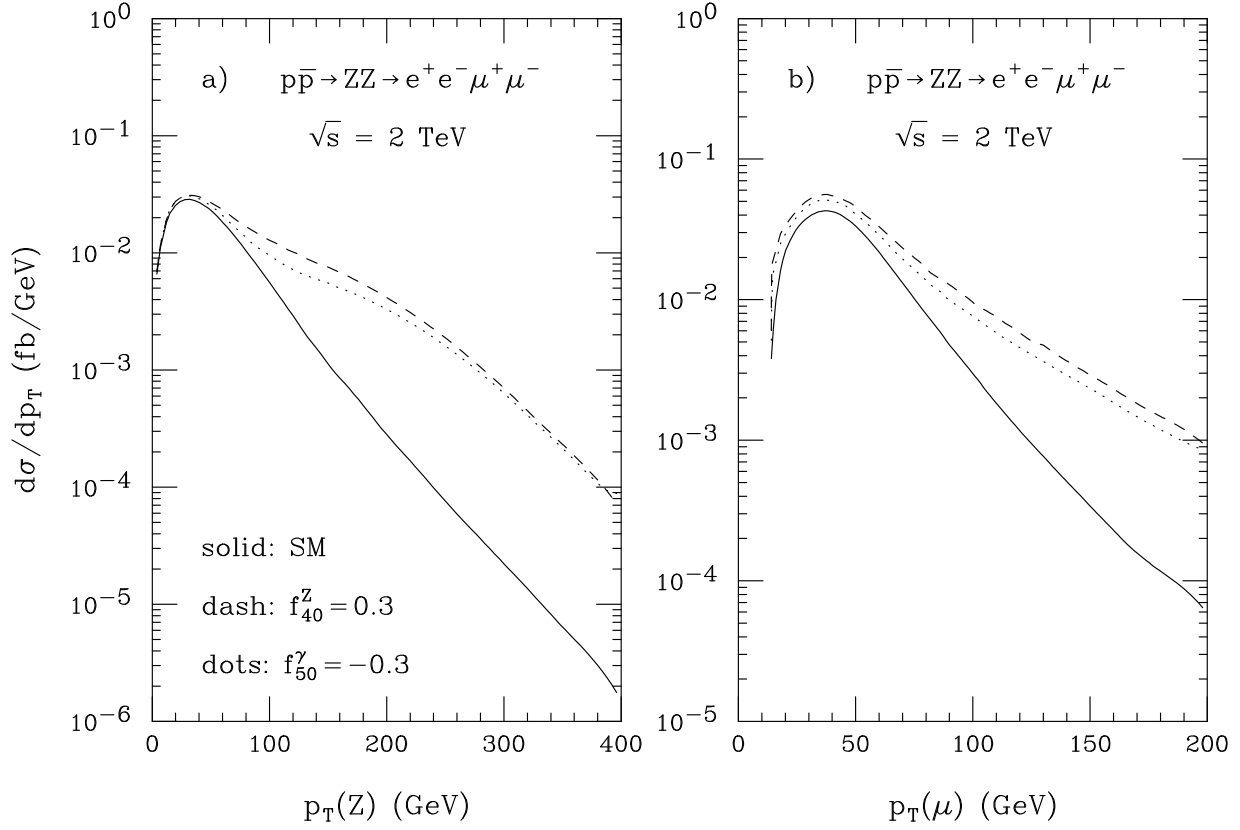}
\caption{The $p_T(Z)$ and $p_T(\mu)$ distributions in 
$p\bar{p}\to ZZ\to e^+e^-\mu^+\mu^-$ at the Tevatron ($\sqrt{s} = 2$~TeV) in 
the SM (solid line), for $f_{40}^Z=0.3$, and $f_{50}^\gamma=-0.3$. The
cuts imposed are described in detail in Sec.~\ref{sec:threeone}. The
form factor scale has been set to $\Lambda_{FF}=750$~GeV.}
\label{fig:pT.4l.tev}
\end{figure}

\begin{figure}
\phantom{x}
\vskip 15cm
\includegraphics{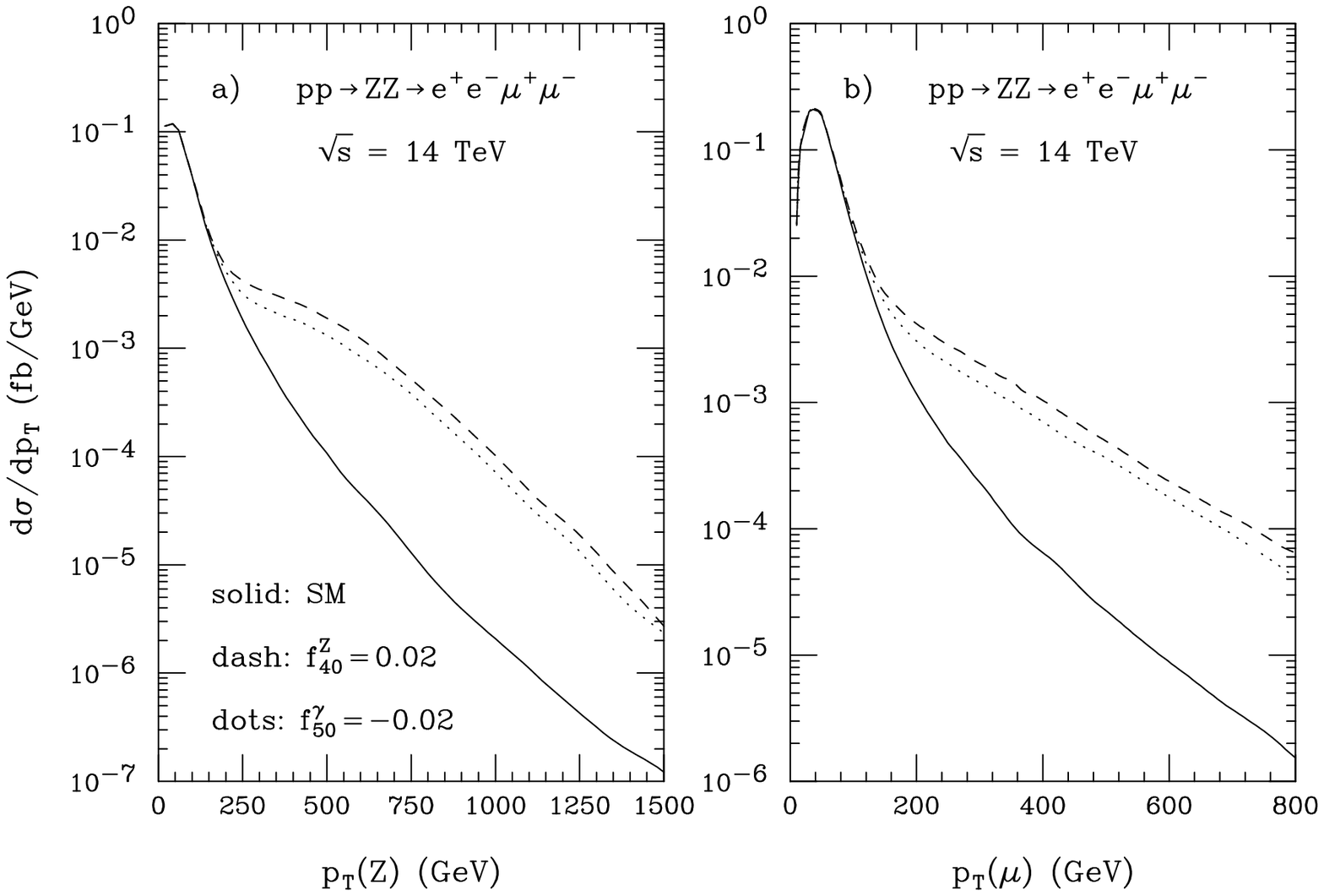}
\caption{The $p_T(Z)$ and $p_T(\mu)$ distributions in 
$pp\to ZZ\to e^+e^-\mu^+\mu^-$ at the LHC ($\sqrt{s} = 14$~TeV) in 
the SM (solid line), for $f_{40}^Z=0.02$, and $f_{50}^\gamma=-0.02$. The
cuts imposed are described in detail in Sec.~\ref{sec:threeone}. The
form factor scale has been set to $\Lambda_{FF}=2$~TeV.}
\label{fig:pT.4l.lhc}
\end{figure}

Similar to the $WWV$ and $Z\gamma V$ couplings~\cite{Wcoupling,Zcoupling}, the 
effects of anomalous $ZZV$ couplings are enhanced at large energies. A typical 
signal of nonstandard $ZZZ$ and $ZZ\gamma$ couplings thus will be a broad 
increase in the $ZZ$ invariant mass distribution, the $Z$ transverse momentum 
distribution and the $p_T$ distribution of the $Z$ decay products. The
$p_T(Z)$ 
and $p_T(\mu)$ distributions for the Tevatron ($p\bar p$ collisions at
$\sqrt{s}=2$~TeV) and the LHC are shown in Figs.~\ref{fig:pT.4l.tev}
and~\ref{fig:pT.4l.lhc}, respectively. Results are shown for the SM and
two $ZZV$ couplings, $f_{40}^Z = 0.3~(0.02)$ and 
$f_{50}^\gamma = -0.3~(-0.02)$ at the Tevatron (LHC). Here, and in all
subsequent figures, only one $ZZV$ coupling is allowed to be non-zero at 
a time. 

Terms proportional to $f_4^V$ and $f_5^V$ in the matrix elements have
identical  
high energy behavior. Differences in the differential cross sections at high 
energies between $ZZZ$ and $ZZ\gamma$ couplings are thus controlled by the 
$Zf\bar{f}$ and $\gamma f\bar{f}$ couplings, and by the parton distribution 
functions. At the Tevatron these result in differential cross sections which 
differ by only a few percent for $\hat{s}\gg M_Z^2$ if
$|f_{i0}^Z|=|f_{i0}^\gamma|$ 
($i=4,5$). Slightly larger differences are observed at intermediate energies 
and transverse momenta. Since $f^V_4$ violate $CP$ conservation, terms in the 
helicity amplitudes proportional to those couplings do not interfere with the 
SM terms. Cross sections thus are independent of the sign of
$f_4^V$. Interference effects between the anomalous and SM contributions 
to the helicity amplitudes, however, do occur for $f^V_5$. The magnitude of
the interference effects in the transverse momentum and $ZZ$ invariant
mass distributions unfortunately is small. 

While it would be difficult to discriminate between the various $ZZV$
couplings in the $m_{ZZ}$ or $p_T$ distributions, it should be easy to
distinguish between the SM Higgs boson which decays in a pair of $Z$
bosons, $H\to ZZ$, and anomalous $ZZV$ couplings. This will be important
at the LHC, where final states resulting from $pp\to ZZ$ are prime Higgs 
boson search channels. Anomalous
gauge boson couplings lead to a broad increase in the differential cross 
section at large energies and transverse momenta, whereas a scalar resonance
produces a Breit-Wigner resonance in the $m_{ZZ}$ distribution and a
Jacobian peak in the $p_T(Z)$ spectrum. In addition, 
the correlation of the angular distributions of the $Z$ decay
leptons in the $Z$ boson rest frames may be used to discriminate between 
a Higgs boson and $ZZV$ couplings. $Z$ bosons originating from Higgs boson
decay are mostly longitudinally polarized~\cite{higgs}, whereas
anomalous $ZZV$ couplings lead to one transversely polarized and one
longitudinally 
polarized $Z$ boson (see Eqs.~(\ref{eq:A1}) and~(\ref{eq:A2})). Since
the $Z$ boson coupling to charged leptons is almost purely axial vector
like, transversely polarized $Z$ bosons lead to a angular distribution
for the $Z$ decay leptons which is proportional to $(1+\cos^2\theta^*)$, 
where $\theta^*$ is the polar angle in the $Z$ boson rest frame with
respect to the flight direction of the $Z$ boson in the $ZZ$ rest
frame. On the other hand, the angular distribution for longitudinal
$Z$'s is proportional to $\sin^2\theta^*$. 

\begin{figure}
\phantom{x}
\vskip 15cm
\includegraphics{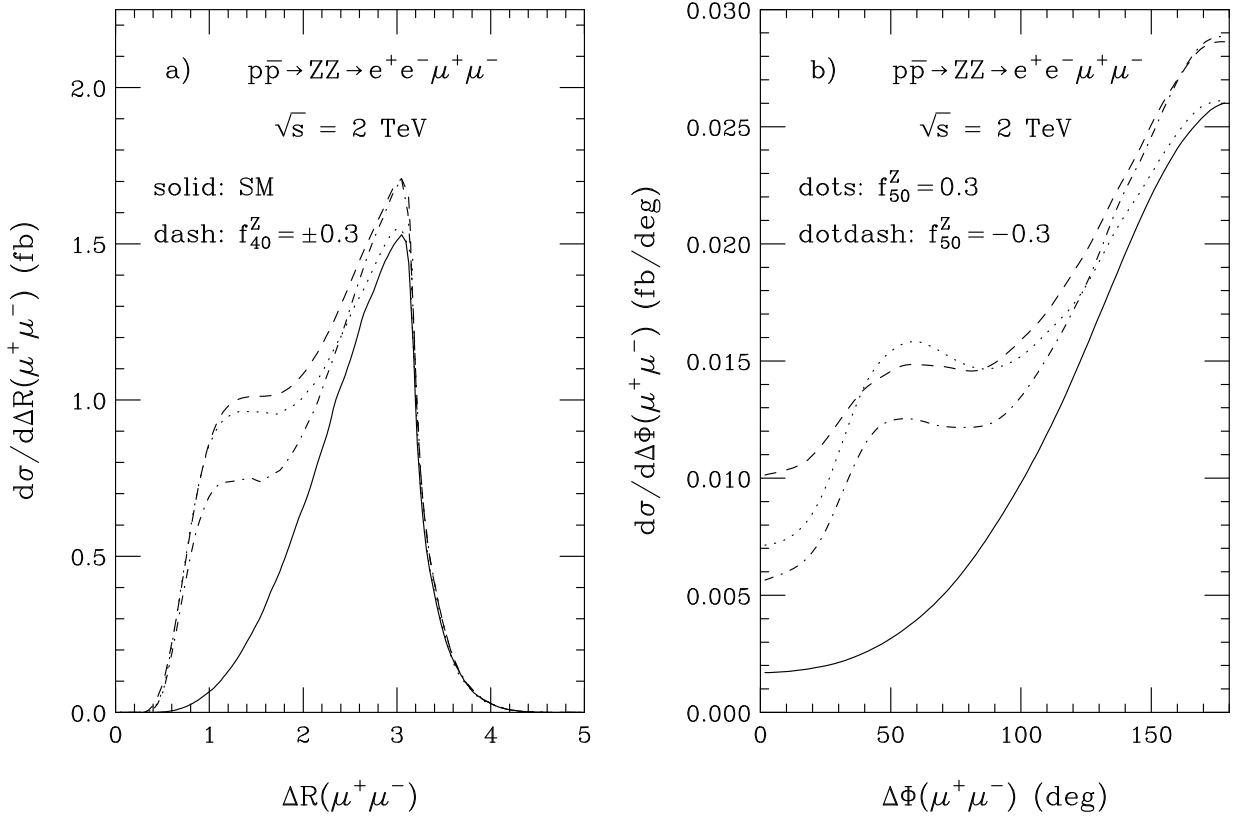}
\caption{The $\Delta R(\mu^+\mu^-)$ and $\Delta\Phi(\mu^+\mu^-)$ distributions 
in $p\bar{p}\to ZZ\to e^+e^-\mu^+\mu^-$ at the Tevatron ($\sqrt{s} = 2$~TeV) 
in the SM and in the presence of non-standard $ZZZ$ couplings. The
cuts imposed are described in detail in Sec.~\ref{sec:threeone}. The
form factor scale has been set to $\Lambda_{FF}=750$~GeV.}
\label{fig:ang.4l.tev}
\end{figure}

In order to distinguish $f_4^V$ and $f_5^V$, and to determine the sign
of $f_5^V$, the $\Delta R$ distribution and the distribution of the
opening angle in the transverse plane, $\Delta\Phi$, of the 
$\ell^+\ell^-$ pair originating from the decay of a $Z$ boson
may be helpful, if deviations from the SM predictions are found in 
the $m_{ZZ}$ and the transverse momentum differential cross sections. 
Figure~\ref{fig:ang.4l.tev} shows the $\Delta R(\mu^+\mu^-)$ and
$\Delta\Phi(\mu^+\mu^-)$ distributions for $ZZ\to e^+e^-\mu^+\mu^-$ at
the Tevatron in the SM and for non-standard $ZZZ$ couplings. Similar 
results are obtained for the corresponding
distributions of the $e^+e^-$ pair, and for the $ZZ\gamma$ couplings 
$f_{4,5}^\gamma$. The SM 
$\Delta R(\mu^+\mu^-)$ and $\Delta\Phi(\mu^+\mu^-)$ differential 
cross sections are dominated by the threshold region, 
$\sqrt{\hat{s}}\approx 2M_Z$, where the $Z$ boson momenta are small and the 
decay leptons tend to be back-to-back, i.e. the distributions are
strongly peaked at $\Delta R\approx 3$ and $\Delta\Phi=180^\circ$. 
Anomalous couplings affect the cross section mostly at large $Z$-boson 
transverse momentum. Due to the Lorentz boost, the relative opening angle 
between the leptons originating from $Z\to\ell^+\ell^-$ decreases with 
increasing $p_T$. The deviations due to non-standard $ZZV$ couplings in the 
$\Delta R(\mu^+\mu^-)$ and $\Delta\Phi(\mu^+\mu^-)$ distributions 
are therefore concentrated at rather small values. 
Figure~\ref{fig:ang.4l.tev} demonstrates that the $\Delta\Phi(\mu^+\mu^-)$
distribution would be particularly useful in separating the individual $ZZV$
couplings. The shapes of the $\Delta\Phi(\mu^+\mu^-)$ distributions for 
$f_{40}^Z=0.3$,
$f_{50}^Z=0.3$ and $f_{50}^Z=-0.3$ differ considerably, and for a
sufficient number of events it should be possible to distinguish $f_4^V$
from $f_5^V$, and to determine the sign of $f_5^V$. Unfortunately, both
the $\Delta R(\mu^+\mu^-)$ and the $\Delta\Phi(\mu^+\mu^-)$ differential 
cross sections 
are useless in distinguishing $f_i^Z$ from $f_i^\gamma$ ($i=4,5$). As for 
the $ZZ$ invariant mass distribution and the transverse momentum
distributions, the differential cross sections for $f_i^Z=f_i^\gamma$
differ very little in shape and magnitude. 

\begin{figure}
\phantom{x}
\vskip 15cm
\includegraphics{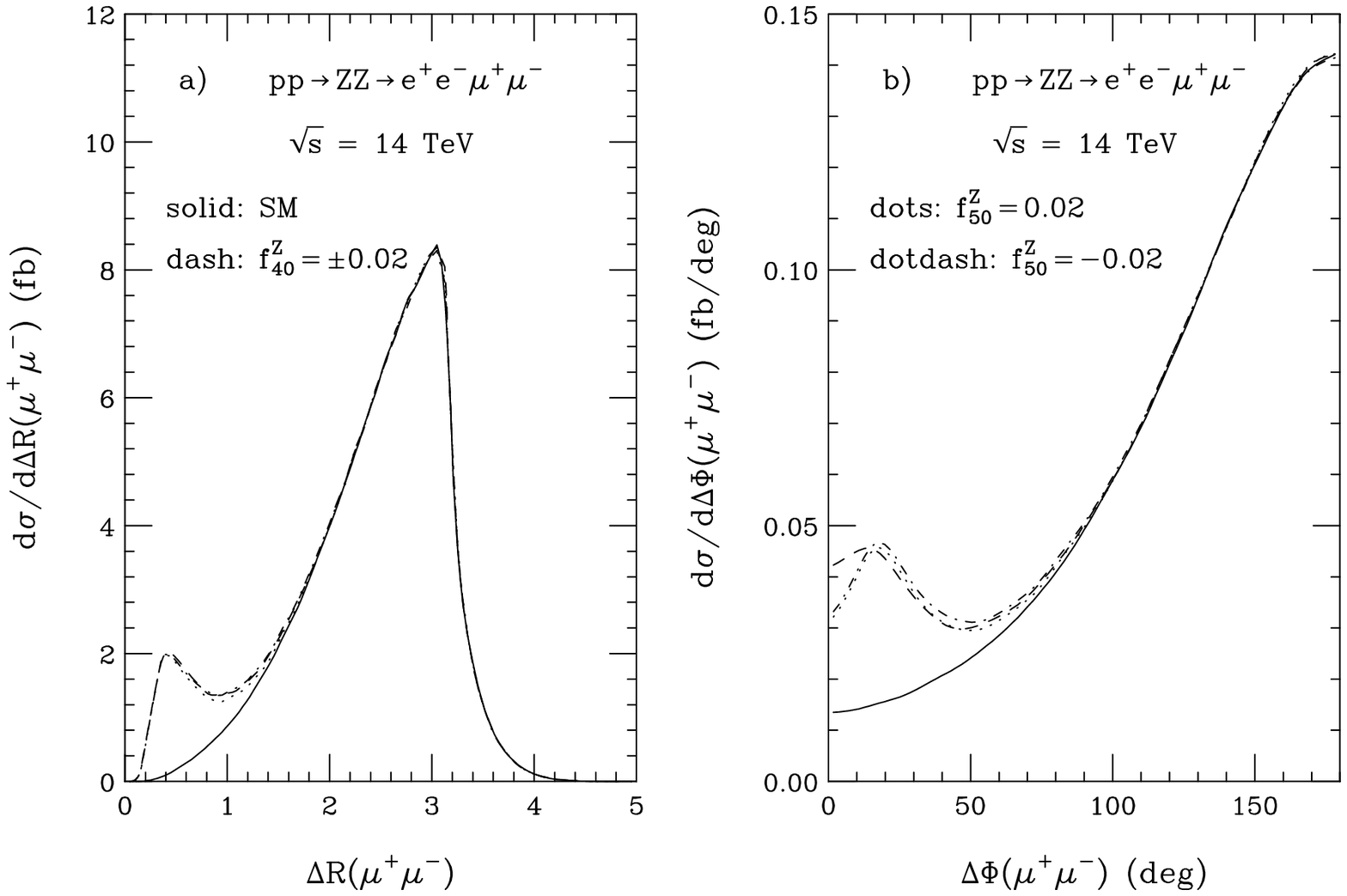}
\caption{The $\Delta R(\mu^+\mu^-)$ and $\Delta\Phi(\mu^+\mu^-)$ distributions 
in $pp\to ZZ\to e^+e^-\mu^+\mu^-$ at the LHC ($\sqrt{s} = 14$~TeV) 
in the SM and in the presence of non-standard $ZZZ$ couplings. The
cuts imposed are described in detail in Sec.~\ref{sec:threeone}. The
form factor scale has been set to $\Lambda_{FF}=2$~TeV.}
\label{fig:ang.4l.lhc}
\end{figure}

At the Tevatron, the small number of $ZZ\to 4$~leptons events will limit 
the usefulness of the $\Delta R(\ell^+\ell^-)$ and
$\Delta\Phi(\ell^+\ell^-)$ distributions. At the LHC, the expected
number of events is much larger; however, the magnitude of the
interference effects between the SM and the anomalous contributions to
the helicity amplitudes is significantly smaller than at the
Tevatron. This can be easily understood from the high energy behavior of 
the anomalous contributions to the differential cross section. The
differential cross section is proportional to the squared amplitude,
which contains the SM terms, terms linear in the anomalous couplings and 
terms which are quadratic in the $ZZV$ couplings. The terms linear in
the anomalous couplings originate from the interference between the SM
amplitude and the anomalous contributions, and are proportional to
$(\sqrt{\hat s}/M_Z)^3$. Terms quadratic in the $ZZV$ couplings on the
other hand are proportional to $(\hat s/M_Z^2)^3$ and thus grow much
faster with $\hat s$. Due to the much higher parton center of mass
energies accessible at the LHC, interference effects thus play a smaller role 
than at the Tevatron. The $\Delta R(\mu^+\mu^-)$ and the
$\Delta\Phi(\mu^+\mu^-)$ distributions at the LHC for the SM,
$f_{40}^Z=0.02$, $f_{50}^Z=0.02$, and $f_{50}^Z=-0.02$ are shown in
Fig.~\ref{fig:ang.4l.lhc}. Both distributions are very insensitive to
the sign of $f_{50}^Z$. The shape of the $\Delta\Phi(\mu^+\mu^-)$ 
distribution differs for $f_{40}^Z$ and $f_{50}^Z$ for 
$\Delta\Phi(\mu^+\mu^-)< 20^\circ$, whereas the $\Delta R(\mu^+\mu^-)$
distributions for $f_{40}^Z$ and $f_{50}^Z$ values of equal magnitude
are almost identical. It would thus be challenging to discriminate
between $f_4^V$ and $f_5^V$, and to determine the sign of $f_5^V$, using
the $\Delta R(\ell^+\ell^-)$ and the $\Delta\Phi(\ell^+\ell^-)$
distributions in the $ZZ\to 4$~leptons mode at the LHC. Since events
in regions where anomalous $ZZV$ couplings have a significant effect 
originate from higher values of $\sqrt{\hat s}$, the typical separation
and the opening angle in the transverse plane are significantly smaller than
at the Tevatron. 


\subsection{The {$\mathbf \ell^+\ell^-\bar\nu\nu$} channel}
\label{sec:nu}

In contrast to the $ZZ\to 4$~leptons mode which is almost background free, 
there are several potentially important background processes if one of
the two $Z$ bosons decays into neutrinos. The advantage of the 
$ZZ\to\ell^+\ell^-\bar\nu\nu$ channel, observable as $\ell^+\ell^-p\llap/_T$, 
is its larger branching fraction. Summing over the three neutrino
species, the raw number of $ZZ\to\ell^+\ell^-\bar\nu\nu$ signal events
before cuts are implemented is about a factor~6 larger than the number
of $ZZ\to 4$~leptons events. 

\begin{figure}
\phantom{x}
\vskip 15cm
\includegraphics{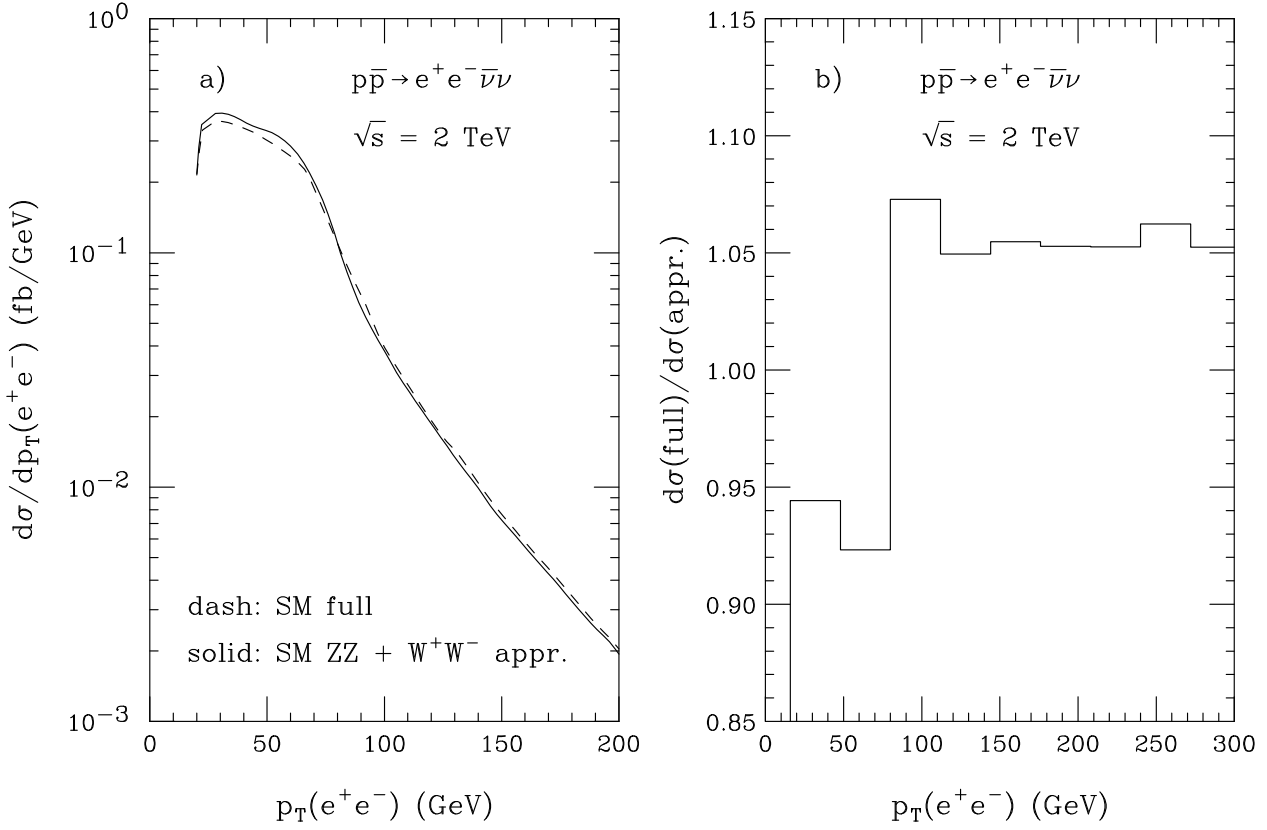}
\caption{Full and approximate results for the transverse momentum
distribution of the $e^+e^-$ pair in 
$p\bar{p}\to e^+e^-\bar\nu\nu$ at the Tevatron ($\sqrt{s} =
2$~TeV) in the SM. The individual distributions obtained using the full
set of contributing Feynman diagrams (solid line) and the subset of
diagrams for $q\bar q\to ZZ\to e^+e^-\bar\nu\nu$ and $q\bar q\to
W^+W^-\to e^+e^-\bar\nu_e\nu_e$ in the double pole approximation (dashed 
line) are shown in part a). Part b) displays the ratio of the full and
approximate differential cross sections. The
cuts imposed are described in detail in the text.} 
\label{fig:ptz.tev.nu.comp}
\end{figure}

Calculation of the signal is similar to that for the four lepton
channel, with the change of one $Z$ coupling from $g_{Z\ell\ell}$ to
$g_{Z\nu\nu}$. There is no photon interference for the $Z$ that decays
to neutrinos. Since it is not possible to restrict the invariant mass of 
the neutrino pair to be in the vicinity of the $Z$ resonance, one has to 
verify that the non-resonant Feynman diagrams ignored in our
calculation do not significantly change 
the differential cross section. There are 10~Feynman diagrams
contributing to $q\bar q\to\ell^+\ell^-\bar\nu_{\ell'}\nu_{\ell'}$,
$\ell\neq \ell'$, and 19 contributing to 
$q\bar q\to\ell^+\ell^-\bar\nu_\ell\nu_\ell$ in the SM. The graphs of 
the $q\bar q\to W^+W^-\to\ell^+\ell^-\bar\nu_\ell\nu_\ell$ background, 
which is described in more detail below, are contained in this set. 
Using {\sc madgraph}~\cite{LS} and the {\sc helas}~\cite{helas} library,
we have calculated the SM cross section for $p\bar p\to e^+e^-
\bar\nu\nu$ including the full set of contributing Feynman diagrams and
summing over the three neutrino species. In 
Fig.~\ref{fig:ptz.tev.nu.comp} we compare the transverse momentum 
distribution of the $e^+e^-$ system for $p\bar p\to e^+e^-\bar\nu\nu$
at the Tevatron resulting from the full set of tree level Feynman 
diagrams with the distribution obtained using the subset of diagrams 
contributing to $q\bar q\to ZZ\to e^+e^-\bar\nu\nu$ and $q\bar q\to 
W^+W^-\to e^+e^-\bar\nu_e\nu_e$ in the double pole approximation. 
Figure~\ref{fig:ptz.tev.nu.comp}a displays the individual $p_T$
distributions, whereas part b) of the figure shows the ratio of the
differential cross sections. In addition to the cuts listed in
Sec.~\ref{sec:threeone} we have imposed a $p\llap/_T>20$~GeV cut in 
Fig.~\ref{fig:ptz.tev.nu.comp}. The non-resonant diagrams are seen to
reduce (enhance) the rate by about $5\%$ for $p_T(e^+e^-)<80$~GeV
($p_T(e^+e^-)>80$~GeV). This is significantly smaller than other
theoretical uncertainties such as the
factorization scale uncertainty in our calculation. In the following
we shall therefore ignore effects of the non-resonant diagrams in channels 
where one of the $Z$ bosons decays into neutrinos.

The most important background processes to $ZZ\to\ell^+\ell^-\bar\nu\nu$ 
production are $t\bar{t}\to W^+W^-b\bar{b}$, 
standard electroweak $W^+W^- +X$ production with 
$W^+W^-\to\ell^+\nu\ell^-\bar\nu$, and $Z(\to\ell^+\ell^-)+1$~jet production 
with the jet rapidity outside the range covered by the detector, thus faking 
missing $p_T$. We will call this last process the ``$Z+1$~jet''
background. We have calculated the $t\bar t$ background using standard
helicity amplitude techniques, fully including the subsequent decays
$t\to Wb$ and $W\to\ell\nu$ and all decay correlations. Finite width 
effects for the top quarks and $W$'s are included. Jets (partons) with
$\Delta R(jj)<0.4$ are merged into a single jet. We do not 
decay the bottom quarks explicitly, but do include a parameterized energy loss 
distribution to make a more realistic simulation of observed final state 
momenta and overall missing momentum. For $W^+W^- +X$
production we make use of the calculation described in
Ref.~\cite{BHO}. For a realistic assessment of the $Z+1$~jet background, a
full-fledged Monte Carlo simulation is required. Here, for a first rough 
estimate, we use a simple parton level calculation. For a jet, 
i.e., a quark or gluon, to be misidentified as $p\llap/_T$ at the
Tevatron (LHC), we shall require that the jet pseudorapidity be
$|\eta(j)|>3.5$ ($|\eta(j)|>5$). 

Additional backgrounds originate from $b\bar b$ production and
$Z\to\tau^+\tau^-$ decays. These backgrounds can be suppressed to a
negligible level by
requiring the angle in the transverse plane between a charged lepton and 
the missing transverse momentum to be between $20^\circ$ and
$160^\circ$, if $p\llap/_T< 50$~GeV. 

Subsequently, in this section we shall focus on the $e^+e^- p\llap/_T$ final
state. Virtually identical results are obtained for $ZZ\to\mu^+\mu^-
p\llap/_T$. For reasons which will become clear shortly, 
we do not include a $K$-factor for the signal cross sections 
in the figures shown in this
section. In addition to the cuts specified in Sec.~\ref{sec:threeone},
we impose a $p\llap/_T>20$~GeV ($p\llap/_T>50$~GeV) cut at the Tevatron
(LHC), and the cut on the angle in the transverse plane between a
charged lepton and the missing transverse momentum discussed above.

\begin{figure}
\phantom{x}
\vskip 15cm
\includegraphics{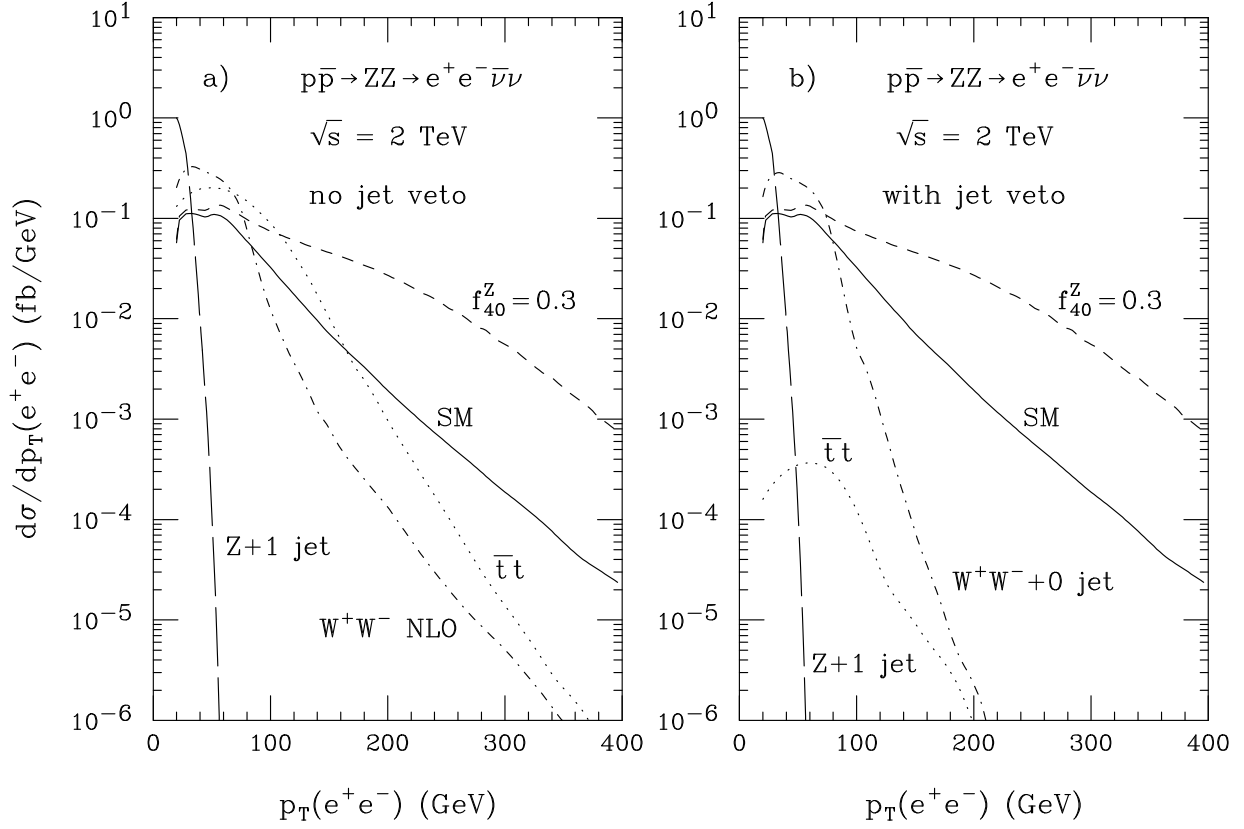}
\caption{Transverse momentum distribution of the $e^+e^-$ pair in 
$p\bar{p}\to ZZ\to e^+e^-\bar\nu\nu$ at the Tevatron ($\sqrt{s} =
2$~TeV) for the SM and for $f_{40}^Z=0.3$, 
together with the differential cross sections from several background
processes (a) without and (b) with a jet veto applied. The
cuts imposed are described in detail in the text. The
form factor scale for nonzero $ZZV$ couplings has been set to
$\Lambda_{FF}=750$~GeV.} 
\label{fig:pT.nu.tev}
\end{figure}

In Fig.~\ref{fig:pT.nu.tev}a we show the transverse momentum
distribution of the $e^+e^-$ pair for $ZZ\to e^+e^- p\llap/_T$ in the SM 
(solid curve) and for $f_{40}^Z=0.3$ (dashed line) at the Tevatron,
together with the differential cross sections of the $t\bar t$ (dotted 
line), $W^+W^-$ (dot-dashed line) 
and $Z+1$~jet (long-dashed line) backgrounds. The cut on the angle in the
transverse plane between a charged lepton and the missing transverse
momentum is responsible for the slight dip in the $ZZ$ differential
cross section curves at $p_T(e^+e^-)\approx 50$~GeV in 
Fig.~\ref{fig:pT.nu.tev}a. 

One observes that all backgrounds shown in Fig.~\ref{fig:pT.nu.tev}a 
are significantly larger than the $ZZ$ signal at 
small values of $p_T(e^+e^-)$. Because of kinematical constraints,
however, the $Z+1$~jet background drops very rapidly with $p_T$. In a more
complete treatment in which soft gluon and/or quark radiation and
hadronization effects are included, one expects that the $p_T(e^+e^-)$
distribution will be somewhat harder for the $Z+1$~jet background,
especially at high transverse momentum. The $Z+1$~jet background
sensitively depends on the rapidity cut above which a jet is assumed to
be misidentified as $p\llap/_T$. Our
assumption that jets with $|\eta(j)|>3.5$ will fake missing transverse
momentum is probably conservative, and the $Z+1$~jet background may well 
be significantly lower than shown in Fig.~\ref{fig:pT.nu.tev}.

The $W^+W^-$ background exceeds
the $ZZ$ signal cross section for $p_T(e^+e^-)<80$~GeV. Since the tail
of the $p_T(e^+e^-)$ distribution for $W^+W^-$ production is very
sensitive to NLL QCD corrections, we show the NLL $p_T(e^+e^-)$
distribution in Fig.~\ref{fig:pT.nu.tev}a. The differential cross
section of the $t\bar{t}$ background is larger than the SM signal for 
$e^+e^-$ transverse momenta as large as 200~GeV, and may thus reduce the
sensitivity to anomalous $ZZV$ couplings. We 
do not show the distributions for the $b\bar{b}$ or $Z\to\tau^+\tau^-$
backgrounds, as they are negligible after the aforementioned angular cut.

The $t\bar t$ background can be virtually eliminated by requiring that
no jets with $p_T(j) > 20$~GeV and $|\eta(j)| < 3.5$ are present. Such a 
jet veto also reduces the $W^+W^-+X$ background at large transverse
momenta. This is shown in Fig.~\ref{fig:pT.nu.tev}b. The remaining
$W^+W^-$ background will only marginally affect the sensitivity to $ZZV$ 
couplings. The jet veto also significantly reduces the size of the NLL
QCD corrections for the $ZZ$ signal, justifying our procedure of not
including a $K$-factor in the signal cross section in this section.

\begin{figure}
\phantom{x}
\vskip 15cm
\includegraphics{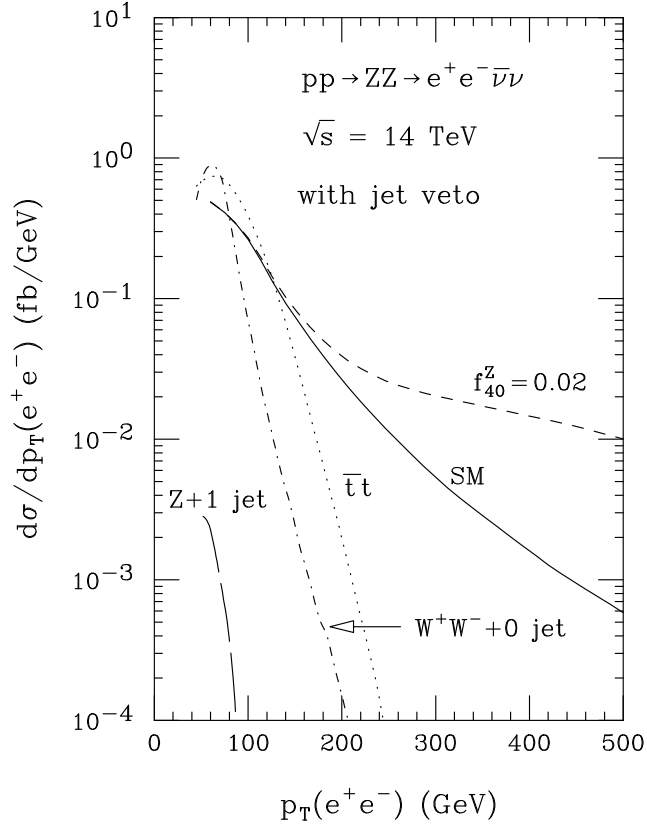}
\caption{Transverse momentum distribution of the $e^+e^-$ pair in 
$pp\to ZZ\to e^+e^-\bar\nu\nu$ at the LHC ($\sqrt{s} = 14$~TeV) for the
SM and for $f_{40}^Z=0.02$, 
together with the differential cross sections from several background
processes. The cuts imposed are described in detail in the text. The
form factor scale for nonzero $ZZV$ couplings has been set to
$\Lambda_{FF}=2$~TeV.} 
\label{fig:pT.nu.lhc}
\end{figure}

The $p_T(e^+e^-)$ distributions for signal and background processes at
the LHC, requiring that no jets
with $p_T(j)>50$~GeV and $|\eta(j)|<5$ are present, are shown in
Fig.~\ref{fig:pT.nu.lhc}. Since the $t\bar t$ production cross section
is more than two orders of magnitude larger than at the
Tevatron, the $t\bar t$ background is significant even when a jet veto
is required. For the cuts imposed, the $t\bar t$ background (dotted 
line) exceeds the signal
for $p_T(e^+e^-)<140$~GeV but is negligible for large transverse
momenta. Both the $W^+W^-$ and $t\bar t$ backgrounds could be reduced 
somewhat by choosing a smaller $e^+e^-$ invariant mass window. Due to 
the more severe missing transverse momentum cut and the
improved rapidity coverage of the hadronic calorimeters of the LHC
experiments, the $Z+1$~jet background is very small. 

\begin{figure}
\phantom{x}
\vskip 15cm
\includegraphics{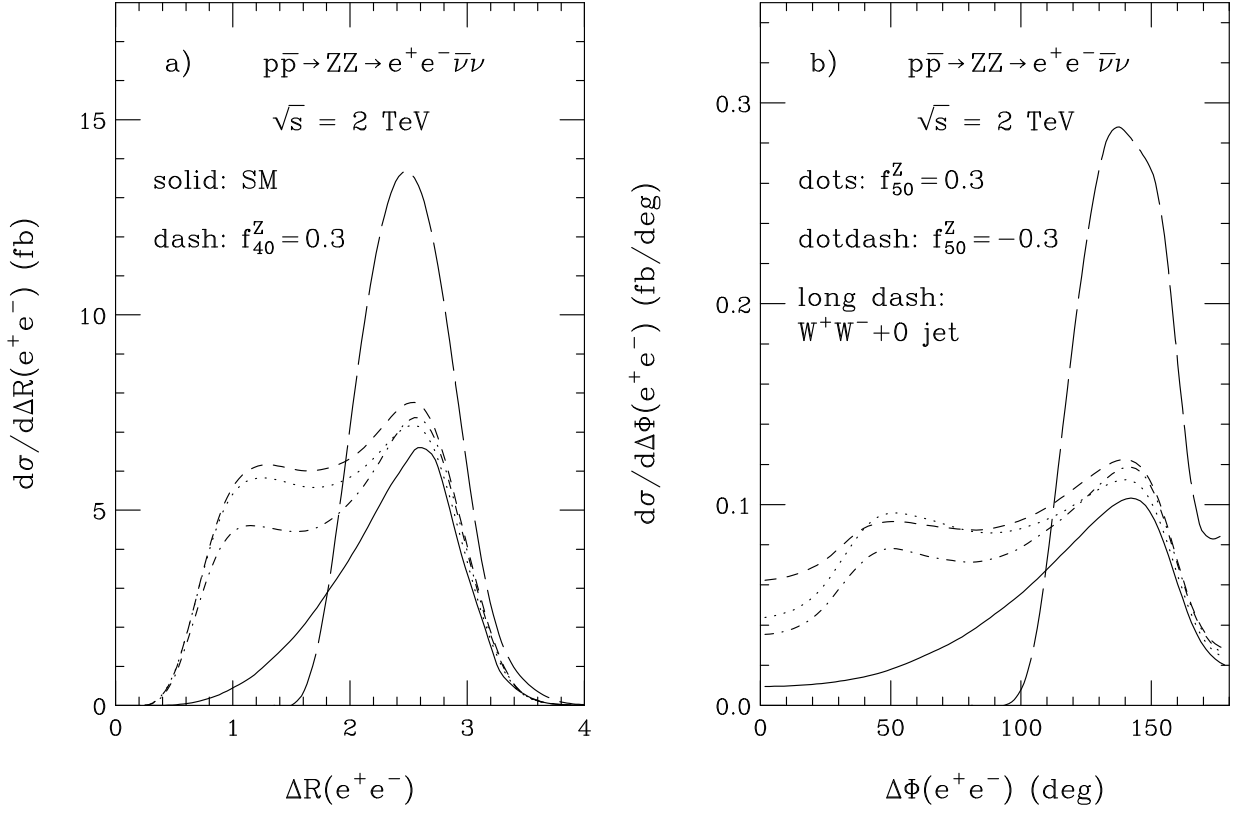}
\caption{The $\Delta R(e^+e^-)$ and $\Delta\Phi(e^+e^-)$ distributions 
in $p\bar{p}\to ZZ\to e^+e^-\bar\nu\nu$ at the Tevatron ($\sqrt{s} = 2$~TeV) 
in the SM and in the presence of non-standard $ZZZ$ couplings. The
cuts described in Sec.~\ref{sec:threeone} and a $p\llap/_T>20$~GeV cut
are imposed. In addition, we require that no jets with $p_T(j)>20$~GeV
and $|\eta(j)|<3.5$ are present. The
form factor scale has been set to $\Lambda_{FF}=750$~GeV.}
\label{fig:ang.nu.tev}
\end{figure}

As in the $ZZ\to 4$~leptons channel, the $\Delta R(e^+e^-)$ and
$\Delta\Phi(e^+e^-)$ distributions would be useful at the Tevatron in 
distinguishing between $f_4^V$ and $f_5^V$, and in determining the sign 
of $f_5^V$. As shown in
Fig.~\ref{fig:ang.nu.tev}, the $W^+W^-$ background is negligible in
those regions of $\Delta R(e^+e^-)$ and $\Delta\Phi(e^+e^-)$ where the
contributions from anomalous $ZZV$ couplings are most pronounced. The 
jet veto imposed in Fig.~\ref{fig:ang.nu.tev} renders the $t\bar t$
background negligible (see Fig.~\ref{fig:pT.nu.tev}b). 
The $p\llap/_T$ cut removes
events where the $Z$ bosons are produced right at threshold and thus 
causes the peak in the $\Delta R(e^+e^-)$ ($\Delta\Phi(e^+e^-)$) distribution 
to shift from $\approx 3$ ($180^\circ$) to $\approx 2.6$ ($\approx
140^\circ$). It has a similar effect on the $W^+W^-+0$~jet
background. Note that the $W$ pair production background vanishes for 
$\Delta R(e^+e^-)<1.4$ and $\Delta\Phi(e^+e^-)<90^\circ$. The $Z+1$~jet
background is not shown in Fig.~\ref{fig:ang.nu.tev} to avoid
overburdening the figure. Qualitatively, the $\Delta R(e^+e^-)$ and
$\Delta\Phi(e^+e^-)$ distributions of the $Z+1$~jet background are
similar to those of the $W^+W^-+0$~jet background. 

\begin{figure}
\phantom{x}
\vskip 15cm
\includegraphics{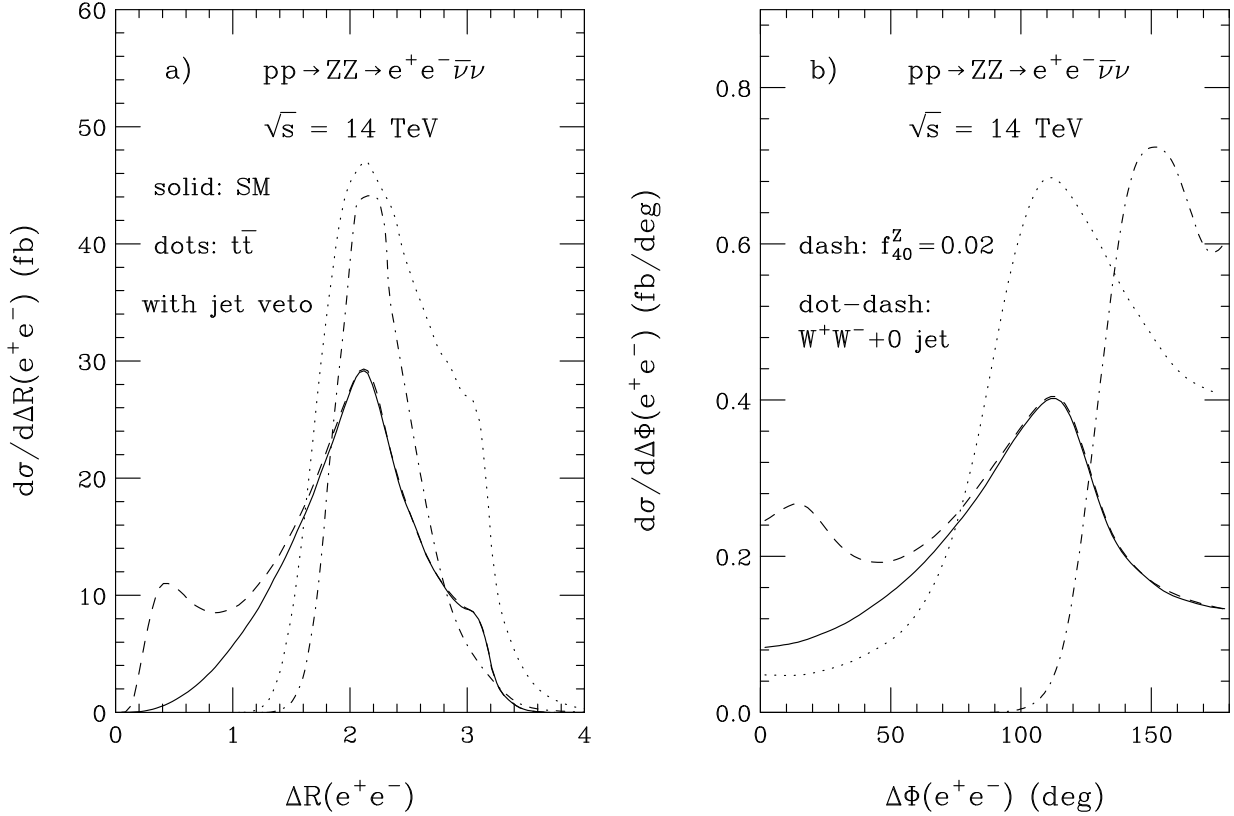}
\caption{The $\Delta R(e^+e^-)$ and $\Delta\Phi(e^+e^-)$ distributions 
in $pp\to ZZ\to e^+e^-\bar\nu\nu$ at the LHC ($\sqrt{s} = 14$~TeV) 
in the SM and for $f_{40}^Z=0.02$ with a form factor scale of
$\Lambda_{FF}=2$~TeV. The dotted and dash-dotted curves represent the
$t\bar t$ and $W^+W^-+0$~jet backgrounds. The
cuts described in Sec.~\ref{sec:threeone} and a $p\llap/_T>50$~GeV cut
are imposed. In addition, we require that no jets with $p_T(j)>50$~GeV
and $|\eta(j)|<5$ are present.}
\label{fig:ang.nu.lhc}
\end{figure}

For completeness, we show the $\Delta R(e^+e^-)$ and
$\Delta\Phi(e^+e^-)$ distributions at the LHC in
Fig.~\ref{fig:ang.nu.lhc}, imposing the same jet veto
requirements as in Fig.~\ref{fig:pT.nu.lhc}. Due to the
higher missing transverse momentum cut, the peak in the SM $ZZ$ 
$\Delta R(e^+e^-)$ ($\Delta\Phi(e^+e^-)$) distribution is shifted to 
$\Delta R(e^+e^-)\approx 2$ ($\Delta\Phi(e^+e^-)\approx 100^\circ$). The 
$\Delta R(e^+e^-)$ distributions of the $t\bar t$ and $W^+W^-+0$~jet
background peak at similar values. Both backgrounds are negligible for 
$\Delta R(e^+e^-)<1.4$. In the SM, the dominant $W^\pm$ helicity at high 
energies in $u\bar u\to W^+W^-$ ($d\bar d\to W^+W^-$) is
$\lambda_{W^\pm}=\mp 1$ ($\lambda_{W^\pm}=\pm 1$)~\cite{whel}. Because
of the $V-A$ nature of the $W\ell\nu$ coupling, the charged leptons in
$W^+W^-\to e^+e^-\bar\nu_e\nu_e$ tend to be emitted either both into
($d\bar d$ annihilation), or both against the flight direction of their
parent $W$ boson ($u\bar u$ annihilation). If a jet veto is imposed, the 
$W^+$ and $W^-$ in $W$ pair production are almost back to back in the
transverse plane, and so are the $W$ decay leptons. The
$\Delta\Phi(e^+e^-)$ distribution of the $W^+W^-+0$~jet background thus
peaks at a significantly larger angle than that of the $ZZ$ signal and
of the $t\bar t$ background (see Fig.~\ref{fig:ang.nu.lhc}b). While the 
$t\bar t$ background is smaller than the SM $ZZ$ signal for
$\Delta\Phi(e^+e^-)<50^\circ$, it is not negligible in this region.


\subsection{The semi-hadronic channels}

The decay modes where one of the two $Z$ bosons decays hadronically have 
much larger branching fractions than the $ZZ\to 4$~leptons and the
$ZZ\to\ell^+\ell^-\bar\nu\nu$ channels, but also much higher
backgrounds. Nevertheless, these channels may be useful in searching for 
$ZZV$ couplings: both CDF~\cite{cdfww} and D\O~\cite{d0ww} have
successfully used the $WW,\,WZ\to\ell\nu jj$ channels to constrain
anomalous $WWV$ couplings in the past. 
 
The main background sources are QCD $Z+2$~jet (``$Zjj$'') production and $WZ$ 
production where the $W$ decays hadronically into a pair of jets. In both
cases the jet pair invariant mass is constrained to be near the $Z$ pole, 
Eq.~(\ref{eq:mass2}). The range of $jj$ invariant masses considered here 
roughly corresponds to $M_Z\pm 2\sigma_{jj}$ where $\sigma_{jj}$ is the
two jet invariant mass resolution of the detector, which typically is
$\sigma_{jj}=5-9$~GeV~\cite{CMS-ATLAS,phot}. The $Z$ boson decays to 
either $e^+e^-,\,\mu^+\mu^-$ or invisibly to neutrinos. 

The $Zjj$ background consists of QCD real-emission corrections to $Z$
production. These subprocesses include~\cite{Kst}
\begin{equation}
q g \to q g \ell^+ \ell^- , \qquad  q q' \to q q' \ell^+ \ell^- \, ,
\nonumber
\end{equation}
which are dominated by $t$-channel gluon exchange, and all crossing-related 
processes, such as
\begin{equation}
q \bar{q} \to g g \ell^+ \ell^- , \qquad g g \to q \bar{q} \ell^+ \ell^- \, .
\nonumber
\end{equation}
Similar to the treatment of the signal processes, we use a
parton-level Monte-Carlo program based on the work of Ref.~\cite{BHOZ_Vjj} 
to model the QCD $Zjj$ background. 
All interference effects between virtual photon and $Z$-exchange are included 
for charged lepton final states; for final state neutrino pairs there is no 
photon contribution. $\alpha_s$ running at one loop order is 
included, correcting the contribution from each phase space point from the 
input value of $\alpha_s(M_Z)=0.118$.
To compute the $WZ$ background we make use of the calculation presented
in Ref.~\cite{BHOZ_VVnj}. 

Other potentially dangerous background sources are $t\bar{t}$ and $Wjj$
production. The $Wjj$ background contributes only to the 
$\bar\nu\nu jj$ final state. For $ZZ\to\ell^+\ell^- jj$, the $t\bar
t\to\ell^+\nu_\ell\ell^-\bar\nu_\ell \bar bb$ background can be reduced 
by requiring that the
missing transverse momentum is $p\llap/_T<20$~GeV at the Tevatron, and 
$p\llap/_T<40$~GeV at the LHC. The higher value for the LHC is motivated 
by pile-up effects due to 
the large number of interactions per beam crossing at the LHC design
luminosity of ${\cal L}=10^{34}~{\rm cm^{-2}~s^{-1}}$. Pile-up effects
amplify small momentum imbalance effects due to mismeasurements in the
underlying event structure and can lead to a significant fake missing
transverse momentum. Backgrounds from $W^+W^-jj$ production are
significantly smaller than those from $t\bar t$ production after
imposing a $p\llap/_T$ veto, and are ignored in the following. 

For $ZZ\to \bar\nu\nu jj$, suppression of the $t\bar
t\to\ell^+\nu_\ell\ell^-\bar\nu_\ell \bar bb$ and $W(\to\ell\nu)jj$ 
backgrounds is possible by
requiring that there are no charged leptons with $p_T(\ell)>10$~GeV and
$|\eta(\ell)|<2.5$ present in the event. The $t\bar
t\to\ell\nu jj\bar bb$ background can be further reduced by imposing the
constraint that the event does not contain more than two jets satisfying 
the $p_T$ and pseudo-rapidity cuts described in
Sec.~\ref{sec:threeone}. The $Wjj$ 
background is calculated using the results of Ref.~\cite{BHOZ_Vjj}. 

In the following, for $ZZ\to\ell^+\ell^- jj$, we shall sum over
electrons and muons in the final state and impose a $p\llap/_T<20$~GeV 
($p\llap/_T<40$~GeV) cut at the Tevatron (LHC) in addition to the
charged lepton and jet $p_T$, pseudorapidity and invariant mass cuts
specified in Sec.~\ref{sec:threeone}. For $ZZ\to \bar\nu\nu jj$, we
require a missing transverse momentum of $p\llap/_T>60$~GeV and no charged
leptons with $p_T(\ell)>10$~GeV and $|\eta(\ell)|<2.5$. In addition, the 
number of jets which satisfy the cuts detailed in
Sec.~\ref{sec:threeone}, $n_j$, has to be $n_j=2$. The rather high 
$p\llap/_T$ cut ensures that
backgrounds from heavy quark production and three jet production, where
the rapidity of one of the jets is outside the range covered by the
detector, are sufficiently suppressed. At
the LHC, a $p\llap/_T>60$~GeV cut may well be too low to trigger on
$jjp\llap/_T$ events, especially at high luminosities. However, since
anomalous $ZZV$ couplings lead to deviations only at large transverse
momenta, raising the $p\llap/_T$ cut to 100~GeV or even 200~GeV at the
LHC will have very little impact on the sensitivity of the $ZZ\to \bar\nu\nu
jj$ mode to $ZZV$ couplings. Finally, all signal and background cross 
sections presented in this section are calculated at leading order. 

\begin{figure}
\phantom{x}
\vskip 15cm
\includegraphics{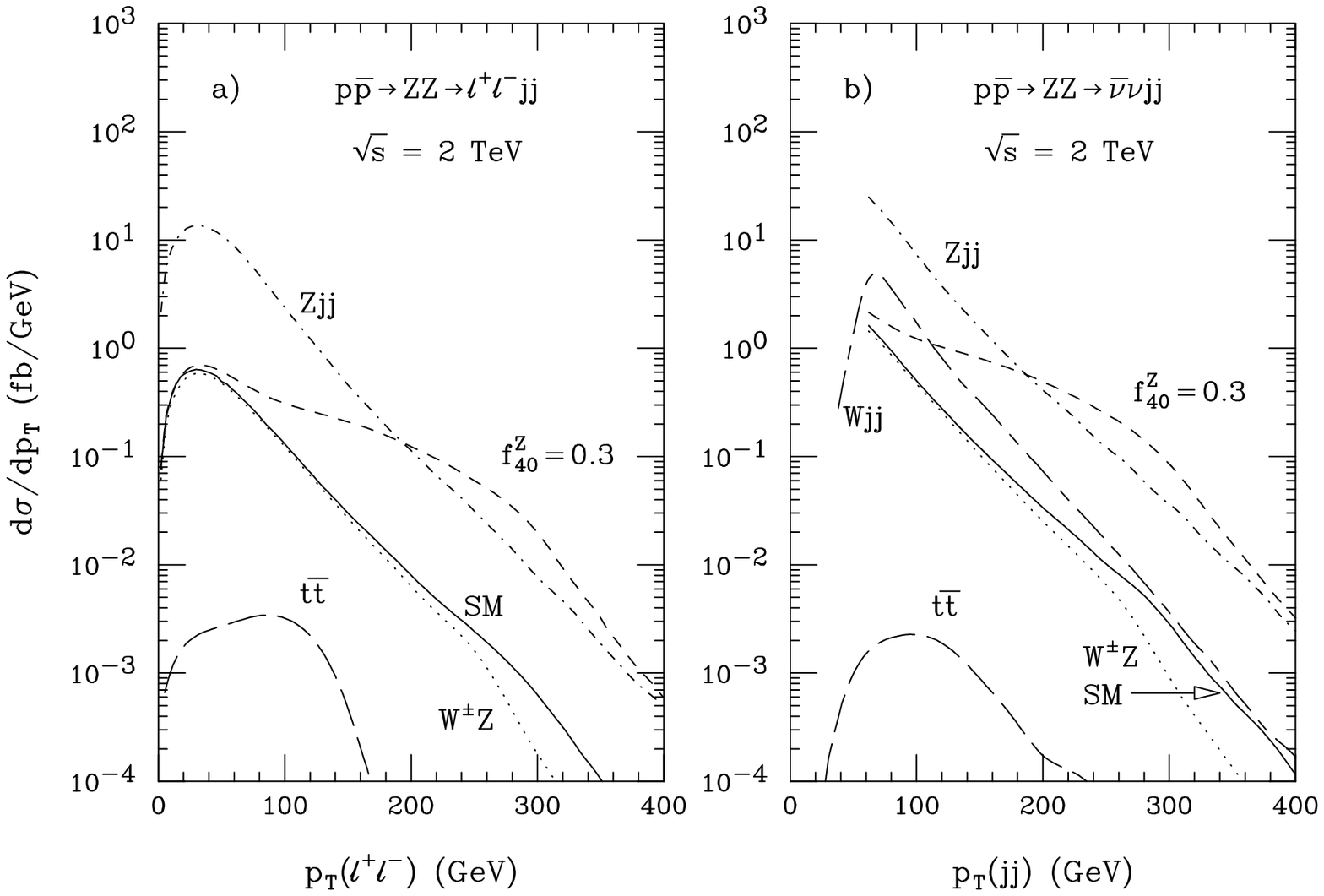}
\caption{Transverse momentum distribution (a) of the $\ell^+\ell^-$ pair in 
$p\bar{p}\to ZZ\to\ell^+\ell^- jj$, and (b) of the jet pair in 
$p\bar{p}\to ZZ\to \bar\nu\nu jj$, at the Tevatron ($\sqrt{s} = 2$~TeV). The SM
prediction is shown together with the cross section for $f_{40}^Z=0.3$
and $\Lambda_{FF}=750$~GeV. Also shown are the differential cross 
sections of various background processes. The cuts imposed 
are described in detail in the text.}
\label{fig:pT.lljj.tev}
\end{figure}

The $p_T(\ell^+\ell^-)$ distribution for $ZZ\to\ell^+\ell^-jj$ is shown 
in Fig.~\ref{fig:pT.lljj.tev}a 
for the Tevatron, while Fig.~\ref{fig:pT.lljj.tev}b illustrates the Tevatron 
$p_T(jj)$ distribution for $\bar\nu\nu jj$ events. We display the SM cross 
section (solid line) together with the main backgrounds. We also show
the $ZZ$ cross section for $f_{40}^Z = 0.3$ (dashed line). The ``kink'' 
in the $WZ$ and $ZZ$ differential cross sections at
$p_T\approx 250$~GeV is due to the 
$\Delta R(jj)>0.6$ cut which becomes effective only at sufficiently high 
transverse momenta. Because of the missing transverse momentum (charged
lepton) veto for $ZZ\to\ell^+\ell^-jj$ ($ZZ\to\bar\nu\nu jj$), the $t\bar 
t$ background is negligible at the Tevatron. The $Wjj$ background, which
contributes only to the $\bar\nu\nu jj$ final state, is considerably larger
than the SM $ZZ$ signal only for small values of $p_T(jj)$, thus it will
not affect the sensitivity to anomalous couplings significantly. 
The $WZ$ differential cross section (dotted line) is very similar 
to that of the SM signal over most of the $p_T$ range considered, whereas
the $Zjj$ background dominates, overwhelming the signal 
even at very high transverse momenta. Its size is uniformly about one 
order of magnitude 
larger than the SM $ZZ$ signal. It will therefore be very difficult to observe 
$ZZ$ production in the semi-hadronic channels, if the SM prediction 
is correct. 
However, for sufficiently large anomalous $ZZV$ couplings, the $ZZ$ cross 
section exceeds the background at large transverse momenta. The
semi-hadronic channels therefore may still be useful 
in obtaining limits on the $ZZV$ couplings at the Tevatron. 

\begin{figure}
\phantom{x}
\vskip 15cm
\includegraphics{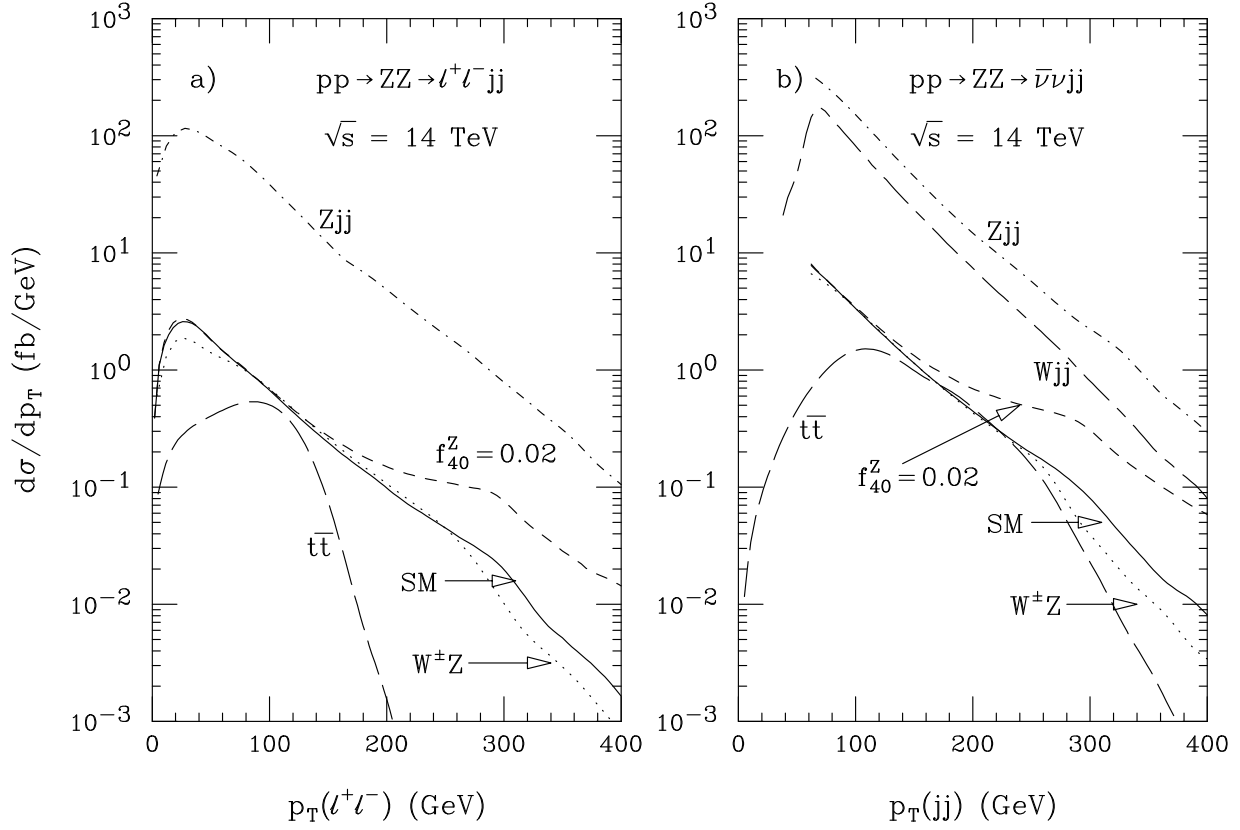}
\caption{Transverse momentum distribution (a) of the $\ell^+\ell^-$ pair in 
$pp\to ZZ\to\ell^+\ell^- jj$, and (b) of the jet pair in 
$pp\to ZZ\to \bar\nu\nu jj$, at the LHC ($\sqrt{s} = 14$~TeV). The SM
prediction is shown together with the cross section for $f_{40}^Z=0.02$
and $\Lambda_{FF}=2$~TeV. Also shown are the differential cross sections 
of various background processes. The cuts imposed are described in
detail in the text.}
\label{fig:pT.lljj.lhc}
\end{figure}

The case is much worse for the LHC, as shown in Fig.~\ref{fig:pT.lljj.lhc}. 
There the rate for $Zjj$ events is almost two orders of magnitude
greater than the 
SM signal, and approximately one order of magnitude greater than that for $ZZ$ 
events for moderate values of $f^V_{i0}$ ($i=4,5$). The $Wjj$
background is a factor~3 to~10 larger than the SM $ZZ\to\bar\nu\nu jj$
signal. Although the
$t\bar t$ background is significantly larger than at the Tevatron, it
has almost no effect on the sensitivity of the $jjp\llap/_T$ final
state to $ZZV$ couplings. 

It should be noted that NLL QCD
corrections could worsen the signal to background ratio. QCD corrections 
enhance the $ZZ$ signal cross section by about a
factor~1.3~\cite{zznlo}. The full NLL
QCD corrections to $Zjj$ and $Wjj$ production are currently not
known. However, the QCD corrections to $Z\bar bb$ production in the 
limit of massless $b$-quarks increase the cross section by about a
factor~2~\cite{CE}. If the QCD corrections to $Zjj$ and $Wjj$ production 
are of similar size, they will weaken the signal to background ratio by
about a factor~1.5. 

In the $\Delta R$ and $\Delta\Phi$ distributions of the $\ell^+\ell^-$
system and the jet pair, the $Zjj$ background overwhelms the $ZZ$ signal 
even for rather large $ZZV$ couplings. For the semi-hadronic $ZZ$ final
states it will therefore be very difficult to utilize these
distributions in distinguishing the various neutral gauge boson
couplings. 

In our discussion of semi-hadronic final states in $ZZ$ production, we
have considered only inclusive jet rates. With the excellent
$b$-tagging capabilities of the Tevatron~\cite{CDF,dzero} and
LHC~\cite{CMS-ATLAS,cms} experiments, one may
also be able to search for anomalous $ZZV$ couplings using
$ZZ\to\ell^+\ell^- \bar bb$ and $ZZ\to\bar\nu\nu\bar bb$ decays. The 
$ZZ\to\ell^+\ell^- \bar bb$ ($ZZ\to\bar\nu\nu\bar bb$)
cross section is about a factor~10 smaller than the $ZZ\to\ell^+\ell^-
jj$ ($ZZ\to\bar\nu\nu jj$) rate if one requires that both $b$-quarks are 
tagged. The main backgrounds in these channels 
are $Z(\to\ell^+\ell^-)\bar bb$, $Z(\to\bar\nu\nu)\bar bb$ and 
$t\bar t$ production. The
$Z\bar bb$ background is about a factor~3 larger than the
$ZZ\to\ell^+\ell^- \bar bb$ and $ZZ\to\bar\nu\nu\bar bb$ 
signals~\cite{CE}. The signal to background
ratio thus is considerably better than in the $\ell^+\ell^- jj$ and
$\bar\nu\nu jj$ 
cases. However, due to the smaller signal rate, one expects that the
sensitivity limits obtained from $ZZ\to\ell^+\ell^- \bar bb$ 
($ZZ\to\bar\nu\nu\bar bb$ ) will be
about a factor~2 to~3 weaker than those derived from $ZZ\to\ell^+\ell^-
jj$ ($ZZ\to\bar\nu\nu jj$).


\section{Sensitivity Limits}
\label{sec:limits}

In this section we discuss the method to extract sensitivity bounds on 
anomalous couplings, and then determine the bounds on $f^V_4$ and
$f^V_5$ which one expects to achieve with 2~fb$^{-1}$ and 10~fb$^{-1}$
of data at the Tevatron in Run~II, and 10~fb$^{-1}$ and 100~fb$^{-1}$ at
the LHC. For simplicity, we consider only real $ZZV$ couplings.

We calculate $95\%$ confidence level (CL) limits performing a $\chi^2$ test. 
The statistical significance is calculated by splitting the selected $p_T$ 
distribution into a number of bins, each with typically more than five events. 
We use the $p_T(\ell^+\ell^-)$ distribution for all final states except
$\bar\nu\nu jj$ for which we use the $p_T(jj)$ differential cross section.
Other distributions, such as the $ZZ$ invariant mass distribution 
(useful only for $ZZ\to 4$~leptons), or the maximum or minimum transverse 
momenta of the charged leptons or jets, yield similar results. In deriving our 
sensitivity limits, we combine channels with electrons and muons in the final 
state. In each bin the Poisson statistics is approximated by a Gaussian 
distribution. Since we selected bins containing at least five events,
the error introduced by this approximation is very small. The same
method has been used in the past to estimate limits on anomalous
$WW\gamma$ and $Z\gamma V$ couplings for Run~I of the
Tevatron~\cite{Zcoupling,BB1}. The actual limits obtained from experimental
data~\cite{cdfwwv,cdfww,d0wwv,d0ww,cdfzg,d0zg} agree well with the 
predicted bounds. 

In order to derive realistic limits, we allow for a normalization uncertainty 
of $30\%$ of the SM cross section. Backgrounds in the 
$ZZ\to\ell^+\ell^-\bar\nu\nu$, $ZZ\to\ell^+\ell^- jj$ and $ZZ\to\bar\nu\nu jj$ 
channels are included in our calculation. We impose the cuts described
in detail in Sec.~\ref{sec:threeone}. In the $ZZ\to\ell^+\ell^-\bar\nu\nu$ 
case we assume that a jet veto has been imposed to reduce the $t\bar{t}$ 
background and require $p_T(\ell^+\ell^-) > 40$~GeV to eliminate the $Zj$ 
background. For $ZZ\to\ell^+\ell^- jj$ we require that events do not
contain missing transverse momentum of more than 20~GeV (40~GeV) at
the Tevatron (LHC). Finally, for $ZZ\to\bar\nu\nu jj$ we impose a charged 
lepton veto ($p_T(\ell)<10$~GeV if $|\eta(\ell)|<2.5$) and require
exactly two jets to reduce the $Wjj$ and $t\bar t$ backgrounds. 
As before, we use a form factor of the form of Eq.~(\ref{eq:ff}) 
with $n=3$ and $\Lambda_{FF} = 750$~GeV ($\Lambda_{FF} = 2$~TeV) for 
the Tevatron (LHC). 

\begin{figure}
\phantom{x}
\vskip 15cm
\includegraphics{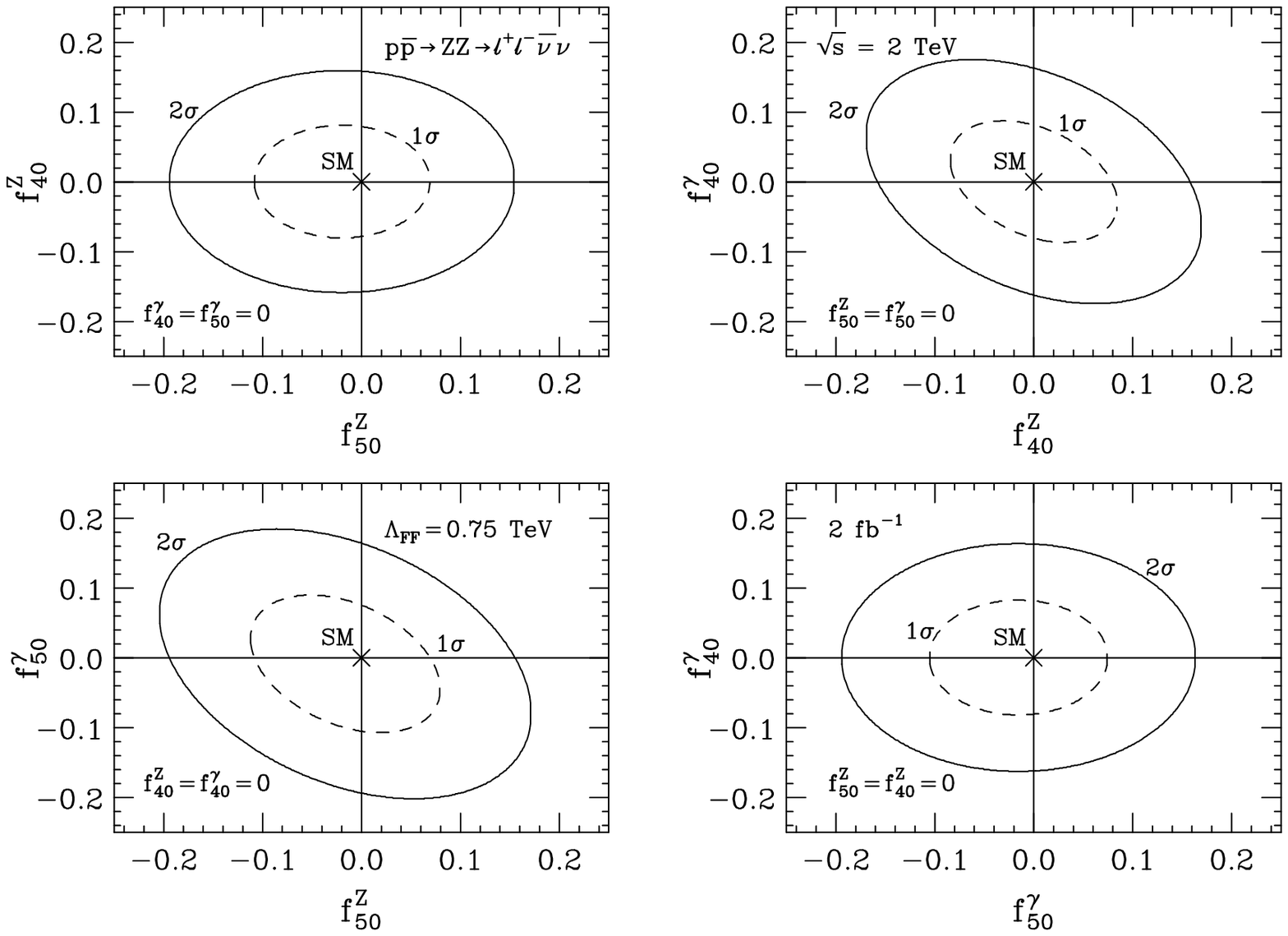}
\vskip 2.cm
\caption{Correlated sensitivity limits for $ZZV$ anomalous couplings in 
$p\bar{p}\to ZZ\to \ell^+\ell^-\bar\nu\nu$ events at the Tevatron Run II 
($\sqrt{s} = 2$~TeV) with 2~fb$^{-1}$ of data. All couplings are assumed
to be real. Shown are $1\sigma$ (dashed lines) and $2\sigma$ (solid
lines) limit contours of all combinations of $f_{i0}^Z$ versus $f_{i0}^\gamma$ 
($i=4,5$). A form factor scale of $\Lambda_{FF}=750$~GeV has been
assumed. In each graph, only those couplings which are plotted against
each other are assumed to be different from their zero SM values.}
\label{fig:lim}
\end{figure}

Non-negligible interference effects between the various $ZZV$ couplings 
are found only between $f_4^Z$ and
$f_4^\gamma$, and between $f_5^Z$ and $f_5^\gamma$. The $f_4^V$ and $f_5^V$
couplings do not interfere, as expected from the CP-odd nature of $f_4^V$
and CP-even nature of $f_5^V$. This result is demonstrated in
Fig.~\ref{fig:lim}, where we show the $1\sigma$ and $2\sigma$ limit
contours  
for $ZZ\to \ell^+\ell^-\bar\nu\nu$ events in 2~fb$^{-1}$ of data at the 
Tevatron Run~II. In each graph, only those couplings plotted against
each other are assumed to be different from their SM values. Plots
similar to those shown in Fig.~\ref{fig:lim} can be obtained for the
LHC, other final states, and different values of $\Lambda_{FF}$. 
As a result of the correlations between $f_4^Z$ and $f_4^\gamma$, and 
between $f_5^Z$ and $f_5^\gamma$, different anomalous coupling 
contributions to the helicity amplitudes may cancel partially, resulting 
in weaker bounds than if only one coupling at a time is allowed to 
deviate from its SM value. 

In Table~\ref{Tevlim} we display $95\%$ confidence level (CL) sensitivity 
limits expected from the Tevatron Run~II for integrated luminosities of 
2~fb$^{-1}$ and 10~fb$^{-1}$, taking into account the correlations 
between different anomalous couplings. Due to the small branching ratio for 
$p\bar{p}\to ZZ\to 4$~leptons, the number of expected events in that channel 
for 2~fb$^{-1}$ is too low to allow for an analysis of the
$p_T(\ell^+\ell^-)$ distribution using the method chosen here. The
bounds obtained from $ZZ\to\ell^+\ell^-\bar\nu\nu$ and $ZZ\to\bar\nu\nu jj$ 
are quite similar. The cross section for $ZZ\to\bar\nu\nu jj$ is about a 
factor~10 larger than that for $ZZ\to\ell^+\ell^-\bar\nu\nu$; however,
the large 
background from $Zjj$ production considerably limits the sensitivity to $ZZV$ 
couplings for $ZZ\to\bar\nu\nu jj$. The limits from the $ZZ\to\ell^+\ell^- jj$ 
and $ZZ\to 4$~leptons channels are about a factor~1.5 and~2 weaker than those 
from $ZZ\to\ell^+\ell^-\bar\nu\nu$ and $ZZ\to\bar\nu\nu jj$. We do not attempt 
to combine limits from different channels. From Table~\ref{Tevlim} it is clear 
that this could result in a significant improvement of the bounds. 

In Table~\ref{LHClim} we display $95\%$ CL sensitivity limits expected from 
the LHC for integrated luminosities of 10~fb$^{-1}$ and 100~fb$^{-1}$. The
most stringent bounds are obtained from the
$ZZ\to\ell^+\ell^-\bar\nu\nu$ channel. The $ZZ\to 4$~leptons channel
yields sensitivity limits which are a factor~1.6 to~2 weaker; while, 
due to the increased $Zjj$ and $Wjj$ backgrounds, the limits which can
be achieved from analyzing the semi-hadronic channels are a factor~2.5
to~4 worse than those found for $ZZ\to\ell^+\ell^-\bar\nu\nu$.

The sensitivity bounds on the $ZZZ$ couplings at the Tevatron (LHC) are 
up to $10\%$ ($20\%$) better than those on the corresponding $ZZ\gamma$ 
couplings. The sensitivities achievable for an integrated
luminosity of 10~fb$^{-1}$ at the Tevatron are about a factor~1.5 to~1.7 
more stringent than those found with 2~fb$^{-1}$. At the LHC, increasing the
integrated luminosity by a factor~10 improves the limits by
approximately a factor~2. These numbers are in good agreement with the
naive scaling law which predicts that sensitivity limits on anomalous
couplings improve roughly with $(\int\!{\cal L}dt)^{1/4}$~\cite{phot}. 
Due to interference with the SM helicity amplitudes, the limits on
$f_{50}^V$ are somewhat sign dependent. In contrast, because of their 
$CP$--violating nature, contributions to the helicity 
amplitudes proportional to $f_{40}^V$ do not interfere with the SM
terms, and the bounds on these couplings do not depend on the sign of
the coupling. 

The $W^+W^-$ and $t\bar t$ backgrounds only marginally influence the
sensitivity limits obtained in the $ZZ\to\ell^+\ell^- \bar\nu\nu$ case. 
The limits derived from $ZZ\to\ell^+\ell^- jj$ and $ZZ\to\bar\nu\nu jj$
production, on the other hand, are significantly degraded by the $Zjj$
and $Wjj$ backgrounds and thus depend on an accurate knowledge of the
signal to background ratio. The signal to background ratio is affected 
by QCD corrections which are not yet fully known, by the factorization
scale uncertainty, and by uncertainties originating from the parton
distribution functions. Changing the signal to background ratio by a
factor~1.5 alters the sensitivity limits by $10-15\%$. 

The sensitivity limits depend significantly on the scale
$\Lambda_{FF}$. For example, at the Tevatron, increasing (decreasing)
the form factor scale to $\Lambda_{FF}=1000$~GeV
($\Lambda_{FF}=500$~GeV) improves (weakens) the bounds which can be 
achieved by a factor~1.5~(2). To a lesser degree, the limits also depend 
on the power $n$ in the form factor, which we have assumed to be
$n=3$. A smaller value of $n$ allows for additional high $p_T$ events
and therefore leads to a somewhat increased sensitivity to the low
energy values of the $ZZV$ couplings. 

Limits on the anomalous couplings depend on the power, $n$, 
and the scale, $\Lambda_{FF}$, of the form factor. These parameters are 
{\it a priori} unknown, as they represent our ignorance of the scale and 
the nature of new 
physics beyond the SM. In Ref.~\cite{glr} it was pointed out that in final 
states without missing transverse momentum one can in principle determine the 
form factor by fitting the $\sqrt{\hat s}$ distribution simultaneously
to $f^V_{i0}$, $\Lambda_{FF}$ and $n$. However, a study~\cite{lhcewk} carried 
out for $WW\gamma$ couplings demonstrated that the method will not produce 
competitive limits. One expects that a similar result is obtained for 
$ZZV$ couplings. Alternatively, if non-zero 
anomalous couplings are observed, the method may be useful in determining the 
shape of the form factor which provides indirect information on the dynamics 
of the underlying new physics. 


\section{Summary and Conclusions}
\label{sec:concl}

$ZZ$ production in hadronic collisions provides an opportunity to probe
neutral gauge boson self-interactions in a direct way. In this paper we
presented a detailed investigation of how well future experiments at the 
Tevatron and LHC will be able to measure the $ZZV$ couplings. Our
calculation has been carried out in the double pole approximation, using
the most general $ZZV$ vertex function which respects Lorentz and
electromagnetic gauge invariance. If both $Z$ bosons are on-shell, there 
are two independent $ZZZ$ ($f^Z_4$ and $f_5^Z$) and two $ZZ\gamma$
couplings ($f^\gamma_4$ and $f_5^\gamma$). $Z$ decays with full decay
correlations and finite $Z$ width effects are included in our
calculation. Non-resonant Feynman diagrams, except for time-like photon
exchange diagrams, and contributions from $gg\to ZZ$ are not taken into
account. The non-resonant Feynman diagrams change the differential cross 
section by at most $5\%$. The contribution from gluon fusion enhances the
cross section by about $1\%$ ($15\%$) at the Tevatron (LHC). NLL QCD
corrections are approximated in our calculation where appropriate by a
simple $K$-factor. 

$S$-matrix unitarity requires that the $ZZV$ couplings are momentum
dependent form factors. We derived constraints on the low energy values
of the $ZZV$ form factors and the shape of the form factor from a
partial wave analysis of the inelastic $ZZ$ production amplitude in
fermion antifermion annihilation (see Eqs.~(\ref{eq:unilim1})
and~(\ref{eq:unilim2})). 

The effects of anomalous $ZZV$ couplings are enhanced at large
energies. They lead to a broad increase in the $ZZ$ invariant 
mass distribution and in various transverse momentum distributions. We
considered four signatures of $ZZ$ production: $ZZ\to 4$~leptons,
$ZZ\to\ell^+\ell^- \bar\nu\nu$, $ZZ\to\ell^+\ell^- jj$, and
$ZZ\to\bar\nu\nu jj$. The $ZZ\to 4$~leptons channel is almost background 
free but suffers from a small event rate due to the small branching
ratio for a $Z$ pair decaying into four charged leptons. The rate for 
$ZZ\to\ell^+\ell^- \bar\nu\nu$ is approximately a factor six larger than 
that for $ZZ\to 4$~leptons; however, $W^+W^-$ and $t\bar t$ production
constitute non-negligible backgrounds at small values of
$p_T(\ell^+\ell^-)$. The $ZZ\to\ell^+\ell^- jj$ and $ZZ\to\bar\nu\nu jj$ 
modes have larger branching ratios than the $ZZ\to 4$~leptons and 
$ZZ\to\ell^+\ell^- \bar\nu\nu$ channels, but also much higher
backgrounds. The main background source for the semi-hadronic
final states is QCD $Z+2$~jet production. The $Z+2$~jet cross section 
is about a factor~20 (50) larger than the $ZZ$ signal at the Tevatron
(LHC). Nevertheless, we found that these channels may be useful in
searching for $ZZV$ couplings. 

All $ZZV$ couplings have similar high energy behavior. This makes it 
difficult to distinguish the various $ZZV$ couplings in the transverse 
momentum and $ZZ$ invariant mass distributions. We found that in the 
$ZZ\to 4$~leptons and $ZZ\to\ell^+\ell^- \bar\nu\nu$ channels, the 
distributions of the lego plot separation, $\Delta R(\ell^+\ell^-)$, 
and the angle in the transverse plane between the charged leptons, 
$\Delta\Phi(\ell^+\ell^-)$, may be useful to distinguish $f_4^V$ from 
$f_5^V$ and to determine the sign of $f_5^V$. Since $f_4^V$ violate 
$CP$ conservation, terms in the helicity amplitudes proportional to 
$f_4^V$ do not interfere with the SM helicity amplitudes, thus it is 
impossible to determine the sign of $f_4^V$ using the 
$\Delta R(\ell^+\ell^-)$ and $\Delta\Phi(\ell^+\ell^-)$ distributions. 
In the semi-hadronic modes, the $ZZ$ signal is overwhelmed by the $Zjj$ 
background in these distributions, even for rather large anomalous 
couplings. Unfortunately, the two distributions will not be useful in 
distinguishing between $f_i^Z$ and $f_i^\gamma$ as they differ very 
little in shape for $ZZZ$ and $ZZ\gamma$ couplings.

In order to determine the bounds on the $ZZV$ couplings which one can
hope to achieve in future Tevatron and LHC experiments, we have
performed a $\chi^2$ test for the four different final states, using the
$\ell^+\ell^-$ and dijet transverse momentum distributions. At the
Tevatron, with an integrated luminosity of 2~fb$^{-1}$, one will be able 
to measure $f_4^V$ and $f_5^V$ with a precision of $15-20\%$ ($95\%$ CL) in the 
$ZZ\to\ell^+\ell^- \bar\nu\nu$ and $ZZ\to\bar\nu\nu jj$ channels for a
form factor scale of $\Lambda_{FF}=750$~GeV. The
limits obtained from the $ZZ\to 4$~leptons and $ZZ\to\ell^+\ell^- jj$
channels are a about a factor~2 and~1.5 weaker than those which can be
achieved in the other two channels (see Table~\ref{Tevlim}). At the LHC
with 100~fb$^{-1}$, the $ZZV$ couplings can be measured with an accuracy of
$3\times 10^{-3}-4\times 10^{-3}$ ($95\%$ CL) in $pp\to ZZ\to\ell^+\ell^-
\bar\nu\nu$ for $\Lambda_{FF}=2$~TeV. The limits obtained from the other
three channels are a factor~1.6 to~4 weaker (see Table~\ref{LHClim}). 

The sensitivity limits which can be achieved at the Tevatron in Run~II
and the LHC should be compared with the bounds from recent measurements
at LEP2 and expectations for a future linear collider, as well as
predictions from theory. The combined
$95\%$ CL limits from the LEP2 experiments presently are~\cite{clara}
\begin{eqnarray}
-0.66< f_4^Z < 0.68, & \qquad & -0.40< f_4^\gamma < 0.38, \\
-1.06< f_5^Z < 0.69,  & \qquad & -0.89< f_5^\gamma < 0.86.
\end{eqnarray}
It should be noted that the LEP2 limits do not contain any form factor 
effects. For the form and scale 
chosen here, these effects weaken the limits by about $20\%$. In 
Run~II, CDF and D\O\ will be able to improve these bounds by at least a
factor~3 to~6. 
At a $e^+e^-$ Linear Collider (LC) with $\sqrt{s}=500$~GeV and an integrated
luminosity of $\int\!{\cal L}dt=100$~fb$^{-1}$, one expects to measure
the $ZZV$ couplings with a precision of~\cite{rosati} $4\times
10^{-3}-6\times 10^{-3}$ ($95\%$ CL). This is comparable to the bounds
which one hopes to achieve at the LHC. 
One loop corrections in the SM, in supersymmetric models, and in models
with heavy fermions, induce $ZZV$ couplings which are of ${\cal
O}(10^{-4})$ or less. 

In view of our present poor knowledge of the $ZZV$ couplings, the direct 
measurement of these couplings in Run~II and at the LHC will constitute
major progress. However, it will be very difficult to achieve a
precision which will make it possible to test the SM one loop prediction 
for the $ZZV$ couplings.

\acknowledgements

We would like to thank S.~Dasu, D.~Denegri and J.~Womersley for useful 
and stimulating
discussions. One of us (UB) would like to thank the Fermilab Theory
Group, where part of this work was carried out, for its generous
hospitality. This work has been supported in part by Department of
Energy contract No.~DE-AC02-76CH03000 and NSF grant PHY-9970703.


\newpage
%
\widetext
\begin{table}
\caption{Sensitivities achievable at $95\%$ CL for anomalous $ZZV$
couplings in 
$p\bar{p}\to ZZ\to 4$~leptons, $p\bar{p}\to ZZ\to\ell^+\ell^-\bar\nu\nu$, 
$p\bar{p}\to ZZ\to\ell^+\ell^- jj$, and $p\bar{p}\to ZZ\to\bar\nu\nu jj$ 
at the Tevatron ($\sqrt{s} = 2$~TeV) for an integrated luminosity of 
(a) 2~fb$^{-1}$, and (b) 10~fb$^{-1}$. The limits for each coupling
apply for arbitrary values of the other couplings. For the form factor
we use the form of Eq.~(\ref{eq:ff}) with $n=3$ and
$\Lambda_{FF}=750$~GeV. The cuts imposed are discussed in the text.}
\label{Tevlim}
\vspace{2mm}
\begin{tabular}{ccccc}
\multicolumn{5}{c}{(a) $\int\!{\cal L}dt=2$~fb$^{-1}$}\\[1mm]
coupling & $ZZ\to 4$~leptons       & $ZZ\to \ell^+\ell^-\bar\nu\nu$ 
         & $ZZ\to \ell^+\ell^- jj$ & $ZZ\to \bar\nu\nu jj$ \\[1mm]
\tableline 
$f_{40}^Z$      & -- & $\matrix{ +0.169 \crcr\noalign{\vskip -4pt} -0.169}$ 
                     & $\matrix{ +0.219 \crcr\noalign{\vskip -4pt} -0.218 }$ 
                     & $\matrix{ +0.159 \crcr\noalign{\vskip -4pt}
-0.160 }$ \\[4mm] 
$f_{40}^\gamma$ & -- & $\matrix{ +0.175 \crcr\noalign{\vskip -4pt} -0.174 }$ 
                     & $\matrix{ +0.222 \crcr\noalign{\vskip -4pt} -0.221 }$ 
                     & $\matrix{ +0.163 \crcr\noalign{\vskip -4pt}
-0.162 }$ \\[4mm] 
$f_{50}^Z$      & -- & $\matrix{ +0.171 \crcr\noalign{\vskip -4pt} -0.204 }$ 
                     & $\matrix{ +0.220 \crcr\noalign{\vskip -4pt} -0.244 }$ 
                     & $\matrix{ +0.162 \crcr\noalign{\vskip -4pt}
-0.184 }$ \\[4mm] 
$f_{50}^\gamma$ & -- & $\matrix{ +0.184 \crcr\noalign{\vskip -4pt} -0.202 }$ 
                     & $\matrix{ +0.229 \crcr\noalign{\vskip -4pt} -0.241 }$ 
                     & $\matrix{ +0.170 \crcr\noalign{\vskip -4pt}
-0.179 }$ \\[1mm] 
\tableline
\multicolumn{5}{c}{(b) $\int\!{\cal L}dt=10$~fb$^{-1}$} \\[1mm]
coupling & $ZZ\to 4$~leptons       & $ZZ\to \ell^+\ell^-\bar\nu\nu$ 
         & $ZZ\to \ell^+\ell^- jj$ & $ZZ\to \bar\nu\nu jj$ \\[1mm]
\tableline 
$f_{40}^Z$      & $\matrix{ +0.180 \crcr\noalign{\vskip -4pt} -0.179 }$ 
                & $\matrix{ +0.097 \crcr\noalign{\vskip -4pt} -0.097 }$ 
                & $\matrix{ +0.146 \crcr\noalign{\vskip -4pt} -0.145 }$ 
                & $\matrix{ +0.106 \crcr\noalign{\vskip -4pt} -0.106 }$ \\[4mm]
$f_{40}^\gamma$ & $\matrix{ +0.185 \crcr\noalign{\vskip -4pt} -0.185 }$ 
                & $\matrix{ +0.100 \crcr\noalign{\vskip -4pt} -0.099 }$ 
                & $\matrix{ +0.148 \crcr\noalign{\vskip -4pt} -0.147 }$ 
                & $\matrix{ +0.109 \crcr\noalign{\vskip -4pt} -0.108 }$ \\[4mm]
$f_{50}^Z$      & $\matrix{ +0.178 \crcr\noalign{\vskip -4pt} -0.216 }$ 
                & $\matrix{ +0.092 \crcr\noalign{\vskip -4pt} -0.120 }$ 
                & $\matrix{ +0.144 \crcr\noalign{\vskip -4pt} -0.167 }$ 
                & $\matrix{ +0.105 \crcr\noalign{\vskip -4pt} -0.127 }$ \\[4mm]
$f_{50}^\gamma$ & $\matrix{ +0.192 \crcr\noalign{\vskip -4pt} -0.213 }$ 
                & $\matrix{ +0.103 \crcr\noalign{\vskip -4pt} -0.115 }$ 
                & $\matrix{ +0.151 \crcr\noalign{\vskip -4pt} -0.163 }$ 
                & $\matrix{ +0.112 \crcr\noalign{\vskip -4pt} -0.121 }$ \\[1mm]
\end{tabular}
\end{table}
\newpage
\begin{table}
\caption{Sensitivities achievable at $95\%$ CL for anomalous $ZZV$
couplings in $pp\to ZZ\to 4$~leptons, $pp\to ZZ\to\ell^+\ell^-\bar\nu\nu$, 
$pp\to ZZ\to\ell^+\ell^- jj$, and $pp\to ZZ\to\bar\nu\nu jj$ 
at the LHC ($\sqrt{s} = 14$~TeV) for an integrated luminosity of 
(a) 10~fb$^{-1}$, and (b) 100~fb$^{-1}$. The limits for each coupling
apply for arbitrary values of the other couplings. For the form factor
we use the form of Eq.~(\ref{eq:ff}) with $n=3$ and
$\Lambda_{FF}=2$~TeV. The cuts imposed are discussed in the text.}
\label{LHClim}
\vspace{2mm}
\begin{tabular}{ccccc}
\multicolumn{5}{c}{(a) $\int\!{\cal L}dt=10$~fb$^{-1}$} \\[1mm]
coupling & $ZZ\to 4$~leptons       & $ZZ\to \ell^+\ell^-\bar\nu\nu$ 
         & $ZZ\to \ell^+\ell^- jj$ & $ZZ\to \bar\nu\nu jj$ \\[1mm]
\tableline 
$f_{40}^Z$      & $\matrix{ +0.0115 \crcr\noalign{\vskip -2pt} -0.0114 }$ 
                & $\matrix{ +0.0060 \crcr\noalign{\vskip -2pt} -0.0060 }$ 
                & $\matrix{ +0.0230 \crcr\noalign{\vskip -2pt} -0.0228 }$ 
                & $\matrix{ +0.0156 \crcr\noalign{\vskip -2pt} -0.0154
}$ \\[4mm] 
$f_{40}^\gamma$ & $\matrix{ +0.0139 \crcr\noalign{\vskip -2pt} -0.0139 }$ 
                & $\matrix{ +0.0072 \crcr\noalign{\vskip -2pt} -0.0072 }$ 
                & $\matrix{ +0.0274 \crcr\noalign{\vskip -2pt} -0.0274 }$ 
                & $\matrix{ +0.0186 \crcr\noalign{\vskip -2pt} -0.0186
}$ \\[4mm] 
$f_{50}^Z$      & $\matrix{ +0.0119 \crcr\noalign{\vskip -2pt} -0.0113 }$ 
                & $\matrix{ +0.0062 \crcr\noalign{\vskip -2pt} -0.0060 }$ 
                & $\matrix{ +0.0226 \crcr\noalign{\vskip -2pt} -0.0220 }$ 
                & $\matrix{ +0.0160 \crcr\noalign{\vskip -2pt} -0.0158
}$ \\[4mm] 
$f_{50}^\gamma$ & $\matrix{ +0.0137 \crcr\noalign{\vskip -2pt} -0.0145 }$ 
                & $\matrix{ +0.0072 \crcr\noalign{\vskip -2pt} -0.0075 }$ 
                & $\matrix{ +0.0268 \crcr\noalign{\vskip -2pt} -0.0274 }$ 
                & $\matrix{ +0.0188 \crcr\noalign{\vskip -2pt} -0.0190
}$ \\[1mm] 
\tableline
\multicolumn{5}{c}{(b) $\int\!{\cal L}dt=100$~fb$^{-1}$} \\[1mm]
coupling & $ZZ\to 4$~leptons       & $ZZ\to \ell^+\ell^-\bar\nu\nu$ 
         & $ZZ\to \ell^+\ell^- jj$ & $ZZ\to \bar\nu\nu jj$ \\[1mm]
\tableline 
$f_{40}^Z$      & $\matrix{ +0.0052 \crcr\noalign{\vskip -2pt} -0.0051 }$ 
                & $\matrix{ +0.0031 \crcr\noalign{\vskip -2pt} -0.0031 }$ 
                & $\matrix{ +0.0130 \crcr\noalign{\vskip -2pt} -0.0128 }$ 
                & $\matrix{ +0.0088 \crcr\noalign{\vskip -2pt} -0.0086
}$ \\[4mm] 
$f_{40}^\gamma$ & $\matrix{ +0.0062 \crcr\noalign{\vskip -2pt} -0.0062 }$ 
                & $\matrix{ +0.0038 \crcr\noalign{\vskip -2pt} -0.0038 }$ 
                & $\matrix{ +0.0154 \crcr\noalign{\vskip -2pt} -0.0154 }$ 
                & $\matrix{ +0.0104 \crcr\noalign{\vskip -2pt} -0.0104
}$ \\[4mm] 
$f_{50}^Z$      & $\matrix{ +0.0053 \crcr\noalign{\vskip -2pt} -0.0051 }$ 
                & $\matrix{ +0.0032 \crcr\noalign{\vskip -2pt} -0.0031 }$ 
                & $\matrix{ +0.0128 \crcr\noalign{\vskip -2pt} -0.0122 }$ 
                & $\matrix{ +0.0092 \crcr\noalign{\vskip -2pt} -0.0088
}$ \\[4mm] 
$f_{50}^\gamma$ & $\matrix{ +0.0061 \crcr\noalign{\vskip -2pt} -0.0065 }$ 
                & $\matrix{ +0.0037 \crcr\noalign{\vskip -2pt} -0.0039 }$ 
                & $\matrix{ +0.0148 \crcr\noalign{\vskip -2pt} -0.0154 }$ 
                & $\matrix{ +0.0104 \crcr\noalign{\vskip -2pt} -0.0108
}$ \\[1mm] 
\end{tabular}
\end{table}
%
%
%
%
%
%
%
\newpage
%
%

\newpage

%
%

\newpage

%
%

\newpage

%
%

\newpage

%
%

\newpage

%
%
\newpage

%
%

\newpage

%
%

\newpage

%
%

\newpage

%
%

\newpage

%
%

\newpage

%
%

\newpage

%
%


\begin{references}

\bibitem{LEP2}
G.~Bella {\it et al.} (The LEP TGC Working Group), LEPEWWG/TGC/2000-01
(March~2000) and references therein.

\bibitem{cdfwwv}
F.~Abe {\it et al.} (CDF Collaboration), Phys. Rev. Lett. {\bf 74}, 1936
(1995) and Phys. Rev. Lett. {\bf 78}, 4537 (1997).

\bibitem{cdfww}
F.~Abe {\it et al.}  (CDF Collaboration), Phys.\ Rev.\ Lett.\  {\bf 75},
1017 (1995).

\bibitem{d0wwv}
S.~Abachi {\it et al.} (D\O\ Collaboration), Phys. Rev. Lett. {\bf 75},
1034 (1995); Phys. Rev. Lett. {\bf 78}, 3634 (1997) and Phys. Rev. 
Lett. {\bf 75}, 1023 (1995); 
B.~Abbott {\it et al.} (D\O\ Collaboration), Phys. Rev. {\bf D58},
051101 (1998), and Phys. Rev. {\bf D58}, 031102 (1998). 


\bibitem{d0ww}
S.~Abachi {\it et al.} (D\O\ Collaboration), Phys.\ Rev.\ Lett.\  {\bf
77}, 3303 (1996); 
B.~Abbott {\it et al.} (D\O\ Collaboration), Phys.\ Rev.\ Lett.\  {\bf
79}, 1441 (1997); hep-ex/9912033, to appear in Phys. Rev. {\bf D}, and
Phys.\ Rev.\  {\bf D60}, 072002 (1999). 

\bibitem{cdfzg}
F. Abe {\it et al.} (CDF Collaboration), Phys. Rev. Lett. {\bf 74}, 1941
(1995). 

\bibitem{d0zg}
S.~Abachi {\it et al.} (D{\O} Collaboration), Phys. Rev. Lett. {\bf 75},
1028 (1995); 
Phys. Rev. Lett. {\bf 78}, 3640 (1997); B.~Abbott {\it et al.} (D{\O}
Collaboration), Phys. Rev. {\bf D57}, 3817 (1998). 

\bibitem{lathuile}
B.~Ashmanskas, talk given at the ``Les Rencontres de Physique de la 
Vall\'ee d'Aoste'', 26 February --~4~March 2000, La Thuile, Vall\'ee d'Aoste,
Italy.

\bibitem{LHC} 
The LHC Study Group, Design Study of the Large Hadron Collider,
CERN 91-03, 1991; L.~R.~Evans, Proceedings of the ``27th 
International Conference on High 
Energy Physics'', Glasgow, Scotland, July 1994, Vol.~II, p.~1417 and
CERN-AC-95-002 (preprint, June~1995).

\bibitem{fawzi}
F.~Boudjema, N.~Dombey, and M.~Krawczyk, Phys. Lett. {\bf B222}, 507
(1989). 

\bibitem{glr}
G.J.~Gounaris, J.Layssac and F.M.~Renard, Phys. Rev. {\bf D61}, 073013
(2000). 

\bibitem{Wcoupling}
K.~Hagiwara, R.~D.~Peccei, D.~Zeppenfeld and K.~Hikasa,
Nucl.\ Phys.\ {\bf B282}, 253 (1987).

\bibitem{anapole}
A.~Barroso, F.~Boudjema, J.~Cole and N.~Dombey, Z.~Phys. {\bf C28}, 149
(1985); 
F.~Boudjema and C.~Hamzaoui, Phys. Rev. {\bf D43}, 3748 (1991).

\bibitem{renard1}
G.J.~Gounaris, J.Layssac and F.M.~Renard, hep-ph/0003143 (March~2000).

\bibitem{renard2}
G.J.~Gounaris, J.Layssac and F.M.~Renard, hep-ph/0005269 (May~2000).

\bibitem{unitarity}
J.~M.~Cornwall, D.~N.~Levin and G.~Tiktopoulos, 
Phys. Rev. Lett. {\bf 30}, 1268 (1973); Phys. Rev. {\bf D10}, 1145
(1974);
C.~H.~Llewellyn Smith, Phys. Lett. {\bf B46}, 233 (1973);
S.~D.~Joglekar, Ann. of Phys. {\bf 83}, 427 (1974).

\bibitem{Zcoupling}
U.~Baur and E.L.~Berger, Phys. Rev. {\bf D47}, 4889 (1993).

\bibitem{BZ_PLB201}
U.~Baur and D.~Zeppenfeld, Phys. Lett. {\bf B201}, 383 (1988). 

\bibitem{pdg}
D.E.~Groom {\it et al.} (Particle Data Group), Eur.~Phys.~J. {\bf C15},
1 (2000).

\bibitem{chang}
D.~Chang, W.Y.~Keung, and P.B.~Pal, Phys. Rev. {\bf D51}, 1326 (1995). 

\bibitem{hag}
K. Hagiwara and D. Zeppenfeld, Nucl. Phys. {\bf B274}, 1 (1986).

\bibitem{gb}
E.W.N.~Glover and J.J.~van der Bij, Nucl. Phys. {\bf B321}, 561 (1989). 

\bibitem{lepsum}
D.~Abbaneo {\it et al.} (The LEP Electroweak Working Group),
CERN-EP-2000-016 (January~2000).

\bibitem{CTEQ}
H.L.~Lai {\it et al.}, Phys.\ Rev.\ {\bf D55}, 1280 (1997).

\bibitem{CDF}
F.~Abe {\it et al.} (CDF Collaboration), FERMILAB-Pub-96/390-E, 
(report, October~1996).

\bibitem{CMS-ATLAS}
A.~Airapetian
{\it et al.} (ATLAS Collaboration), ``ATLAS Detector and Physics 
Performance Technical Design Report, Vol.~2'', CERN-LHCC-99-15,
(report, May~1999).

\bibitem{cdfz}
F.~Abe {\it et al.}  (CDF Collaboration), Phys.\ Rev.\ Lett.\  {\bf 76}, 
3070 (1996) and Phys.\ Rev.\  {\bf D52}, 2624 (1995).

\bibitem{d0z}
B.~Abbott {\it et al.}  (D\O\ Collaboration), Phys.\ Rev.\  {\bf D60}, 
052003 (1999).

\bibitem{zznlo}
B.~Mele, P.~Nason and G.~Ridolfi, Nucl. Phys. {\bf B357}, 409 (1991); 
J.~Ohnemus and J.~Owens, Phys. Rev. {\bf D43}, 3626 (1991);
L.~Dixon, Z.~Kunszt, and A.~Signer, Phys. Rev. {\bf D60}, 114037 (1999);
J.M.~Campbell and R.K.~Ellis, Phys. Rev. {\bf D60}, 113006 (1999).

\bibitem{higgs}
M.J.~Duncan, Phys. Lett. {\bf B179}, 393 (1986); T.~Matsuura and
J.J.~van der Bij, Z.~Phys. {\bf C51}, 259 (1991).

\bibitem{LS} 
T.~Stelzer and W.F.~Long, Comput. Phys. Commun. {\bf 81}, 357 (1994).

\bibitem{helas}
H.~Murayama, I.~Watanabe and K.~Hagiwara, KEK report KEK-91-1 (January
1992). 

\bibitem{BHO}
U.~Baur, T.~Han and J.~Ohnemus, Phys. Rev. {\bf D53}, 1098 (1996).

\bibitem{whel}
K.~Hagiwara, R.~D.~Peccei, D.~Zeppenfeld and K.~Hikasa,
Nucl.\ Phys.\  {\bf B282}, 253 (1987); 
J.~Stroughair and C.~L.~Bilchak, Z.\ Phys.\  {\bf C23}, 377 (1984);
C.~L.~Bilchak, R.~W.~Brown and J.~D.~Stroughair, Phys.\ Rev.\  {\bf
D29}, 375 (1984). 

\bibitem{phot}
U.~Baur {\it et al.}, hep-ph/0005226, to appear in the Proceedings of
the ``Fermilab Run~II Workshop on QCD and Weak Boson Physics'',
Fermilab, 1999.

\bibitem{Kst}
S.~D.~Ellis, R.~Kleiss, and W.~J.~Stirling,
Phys.\ Lett.\ {\bf B154}, 435 (1985);
R.~Kleiss and W.~J.~Stirling, Nucl.\ Phys.\ {\bf B262}, 235 (1985);
Phys.\ Lett.\ {\bf B180}, 171 (1986);
J.~F.~Gunion, Z.~Kunszt, and M.~Soldate, Phys.\ Lett.\ {\bf B163}, 389 (1985);
Erratum, Phys.\ Lett.\ {\bf B168}, 427 (1986);
J.~F.~Gunion and M.~Soldate, Phys.\ Rev.\ {\bf D34}, 826 (1986);
R.~K.~Ellis and R.~J.~Gonsalves, in  ``Proc.\ of the Workshop on Super High
Energy Physics'', Eugene, Oregon (1985), ed.\ D.~E.~Soper, p.~287.

\bibitem{BHOZ_Vjj}
V.~Barger, T.~Han, J.~Ohnemus and D.~Zeppenfeld,
Phys.\ Rev.\ Lett.\  {\bf 62}, 1971 (1989) and
Phys.\ Rev.\ {\bf D40}, 2888 (1989).

\bibitem{BHOZ_VVnj}
V.~Barger, T.~Han, J.~Ohnemus and D.~Zeppenfeld,
Phys.\ Rev.\ {\bf D41}, 2782 (1990).

\bibitem {CE}
J.~Campbell and R.K.~Ellis, Fermilab-Pub-00/145-T (June~2000).

\bibitem{dzero}
S.~Abachi {\it et al.}, (D\O\ Collaboration), Fermilab-Pub-96/357-E,
1996. 

\bibitem{cms}
M.~Beneke {\it et al.}, hep-ph/0003033, Proceedings of the ``1999 CERN
Workshop on Standard Model Physics (and more) at the LHC'', CERN Yellow
Report CERN-2000-004, eds. G.~Altarelli and M.~Mangano.

\bibitem{BB1}
U.~Baur and E.~L.~Berger, Phys.\ Rev.\  {\bf D41}, 1476 (1990).

\bibitem{lhcewk}
S.~Haywood {\it et al.}, hep-ph/0003275, Proceedings of the ``1999 CERN
Workshop on Standard Model Physics (and more) at the LHC'', CERN Yellow
Report CERN-2000-004, eds. G.~Altarelli and M.~Mangano.

\bibitem{clara}
C.~Matteuzzi, talk given at the ``XXXth International Conference on High 
Energy Physics'', Osaka, Japan, July~27~-- August~2, 2000.

\bibitem{rosati}
S.~Rosati, talk given at the Linear Collider Workshop, Padova, Italy,
May~2000. 

\end{references}
\end{document}